\documentclass[runningheads]{llncs}
\usepackage[T1]{fontenc}
\usepackage{graphicx}
\usepackage{algorithm}
\usepackage{algpseudocode}
\usepackage{float}
\usepackage{placeins}
\usepackage{comment}
\usepackage{graphicx}
\usepackage{float}
\usepackage{amssymb}
\usepackage{booktabs}
\usepackage{array}
\usepackage{hyperref}
\usepackage{amsmath}
\usepackage{color}
\usepackage{subcaption}
\usepackage{multirow}
\usepackage{wrapfig}
\usepackage{subcaption} 
\usepackage{esvect}
\usepackage[margin=1in]{geometry}

\begin{document}
\raggedbottom
\title{CyGym: A Simulation-Based Game-Theoretic Analysis Framework for Cybersecurity}
\titlerunning{CyGym: A Cybersecurity Simulation Environment}
\author{Michael Lanier and Yevgeniy Vorobeychik}
\institute{Washington University, Saint Louis, MO 63130, USA}
\maketitle              % typeset the header of the contribution
\begin{abstract}
We introduce a novel cybersecurity encounter simulator between a network defender and an attacker designed to facilitate game-theoretic modeling and analysis while maintaining many significant features of real cyber defense.
Our simulator, built within the OpenAI Gym framework, incorporates realistic network topologies, vulnerabilities, exploits (including-zero-days), and defensive mechanisms. 
Additionally, we provide a formal simulation-based game-theoretic model of cyberdefense using this simulator, which features a novel approach to modeling zero-days exploits, and a PSRO-style approach for approximately computing equilibria in this game.
We use our simulator and associated game-theoretic framework to analyze the Volt Typhoon advanced persistent threat (APT).
Volt Typhoon represents a sophisticated cyber attack strategy employed by state-sponsored actors, characterized by stealthy, prolonged infiltration and exploitation of network vulnerabilities.
Our experimental results demonstrate the efficacy of game-theoretic strategies in understanding network resilience against APTs and zero-days, such as Volt Typhoon, providing valuable insight into optimal defensive posture and proactive threat mitigation. Code available at \url{https://github.com/Lan131/CyGym}.
%This work offers a robust framework for understanding and countering complex cyber threats in critical infrastructure networks.
\end{abstract}
\section{Introduction}

Cybersecurity is at its core an interaction between a \emph{defender}, who aims to protect their assets from compromise, and an \emph{attacker}, who developed and uses computing resources to subvert target systems to obtain sensitive information or prevent the targets from performing their regular functions.
It has thus long been recognized that game theory is a useful tool in the defender's arsenal to reason about the best security posture.
However, common game-theoretic models for cybersecurity are either too abstract~\cite{resourceallocation} or too simplistic~\cite{962836} to be useful in practice.
On the other hand, a host of simulation and emulation tools emerged, but the goals of these are often not well-aligned with game-theoretic modeling approaches---for example, taking either only the defender's or the attacker's perspective~\cite{cyberranges}, or modeling aspects such as network latency at high resolution that are of secondary importance and could be abstracted in game-theoretic analysis~\cite{ns3-manual}.
These limitations are particularly acute if we are to analyze decision making in the context of advanced persistent threats (APTs) and zero-day attacks.

\begin{figure}[h]
  \centering
  % Scale the entire 2-panel block to 0.8×\textwidth (adjust as needed)
  \resizebox{0.8\textwidth}{!}{%
    \begin{minipage}{\textwidth}
      \centering
      % Panel (a)
      \begin{subfigure}[b]{0.45\textwidth}
        \centering
        \includegraphics[width=\linewidth]{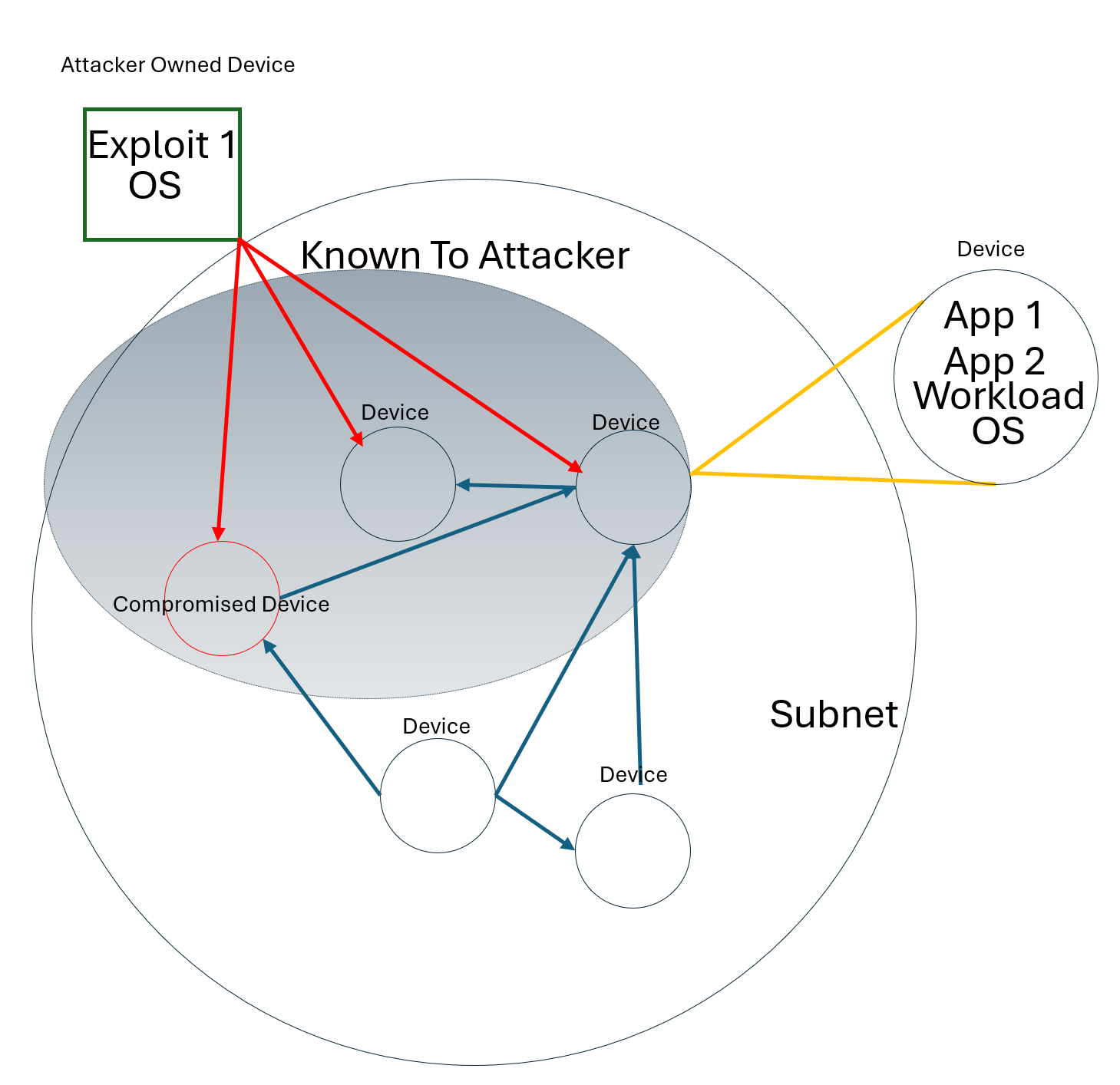}
        \caption{Hierarchical diagram illustrating the simulator’s structure.}
        \label{fig:cyber_sim}
      \end{subfigure}\hfill
      % Panel (b)
      \begin{subfigure}[b]{0.45\textwidth}
        \centering
        \includegraphics[width=\linewidth]{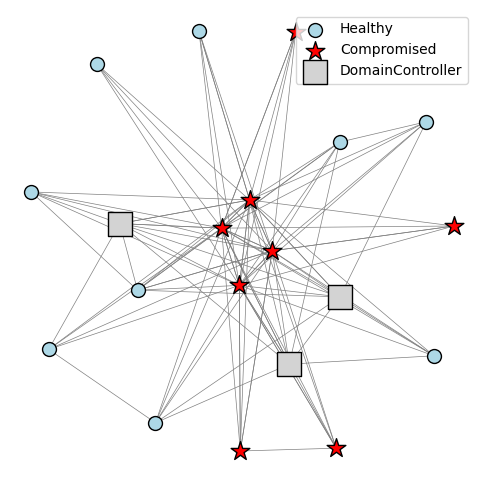}
        \caption{Example underlying graph structure of a subnet.}
        \label{fig:network_viz}
      \end{subfigure}
    \end{minipage}%
  }
  \caption{Illustration of the simulator (a) and an example of a graph structure of the inter-device network (b).}
  \label{fig:combined_viz_and_arch}
\end{figure}

We propose a novel simulation-based game-theoretic framework for cybersecurity that combines a simulation model at intermediate granularity that is focused most on capturing the complexity of the strategic landscape (Figure~\ref{fig:combined_viz_and_arch}) with a formal model as a partially observable stochastic game (POSG).
Moreover, we provide a solution technique that extends the PSRO and double-oracle frameworks to provide for better-response approaches that tackle the combinatorial nature of the action spaces of both players.
A particularly novel aspect of our model is an explicit representation of zero-day attacks in the language of asymmetric information in which a common distribution over \emph{all possible exploits} is shared by both players, but only the attacker knows the actual (zero-day) exploits they can deploy.

We instantiate our simulation-based game theoretic framework to study the attacker-defender interaction dynamics in the context of Volt Typhoon, a recent advanced persistent threat (APT).
According to a joint advisory from the Cybersecurity and Infrastructure Security Agency (CISA), the National Security Agency (NSA), and the Federal Bureau of Investigation (FBI), Volt Typhoon has been actively compromising critical U.S. infrastructure since at least mid-2021 \cite{cisa2024volt,tsa2024cybersecurity}.
Our Volt Typhoon case study first shows that the equilibrium strategies of both players tend to outperform simple heuristic baselines in this case.
Moreover, we qualitatively study the impact of several environment parameters, such as the relative importance of productive workflows to security costs, as well as the availability of zero-days in the attacker's arsenal, on defense and the rate of compromised devices.

To address these limitations, we propose a novel game-theoretic approach that integrates reinforcement learning to model and mitigate the threats posed by Volt Typhoon. Our contributions are threefold:

\begin{enumerate}
    \item \textbf{CyGym: A Gym and Simulator Environment Utilizing NIST Data}: We have developed a custom simulation environment\footnote{\url{https://github.com/Lan131/CyGym}}, built on the OpenAI Gym framework, which leverages real-world data from the National Institute of Standards and Technology (NIST). This environment allows for the realistic modeling of network scenarios, incorporating various vulnerabilities, exploits, and defensive strategies. By using NIST data, we ensure that our simulations are grounded in actual threat landscapes and vulnerability profiles.

    \item \textbf{A Simulation-Based Game-Theoretic Model based on CyGym}: We offer a formal game-theoretic model of cybersecurity encounters that uses the strategic primitives available in CyGym within a partially-observable stochastic game formalism. Our model captures zero-day attacks through the mechanism of asymmetric information about attacker's available strategic options.

    \item \textbf{Double Oracle Solution for Non-Zero Sum Partially Observed Stochastic Games}: We provide an implementation of Double Oracle using reinforcement learning to solve for best response strategies. This solution is scalable to large networks and to the best of our knowledge this is the first time double oracle has been employed in this setting.
    
    \item \textbf{Case Study of Volt Typhoon}: We provide a specific instance of the simulator specific to Volt Typhoon, allowing individual agencies and entities to apply parameters of their organization exposure to such an attack. This case study illustrates how the simulator can be tailored to specific organizational contexts, helping security teams to better understand potential attack vectors and develop more effective defense mechanisms. 
\end{enumerate}

%By modeling the behavior of sophisticated attackers, we offer a valuable tool for proactive threat assessment and mitigation, enhancing the overall resilience of critical infrastructure networks. Entities are encouraged to apply their own Cyber Threat Intelligence \cite{cisa_threat_feeds} data into the parameters of the model to understand their exposure.  
%See Figure~\ref{fig:cyber_sim} for a visual representation.

\section{Related Work}

\subsection{Existing Cybersecurity Simulation Tools}

Existing cybersecurity simulators, such as \cite{gardner2019camouflaged,vartiainen2024development,standen2021cyborg}, have made significant contributions to the field.
\cite{gardner2019camouflaged} introduced camouflaged cyber simulations to ensure experimental validity, emphasizing internal, external, and construct validity.
This simulator is not broadly available or open sourced and relies on emulation technology, making its simulations tailor made to very specific infrastructures.

\cite{vartiainen2024development} created a cybersecurity simulator-based platform for the protection of critical infrastructures.
This solution is used in tandem with a digital twin of a real-world physical system, rendering it dependent on such digital twins and hence difficult for general-purpose analysis.
Because this system is hardware-in-the-loop, it relies on hardware-specific parameters (including latency); changes can require a complete rebuild of the twin.
\emph{CyGym} instead incorporates device latency and busyness as parameters that the developer can readily adjust, avoiding such extensive reconfiguration.

Other tools such as \cite{holm2015cyber,simulacra2019automated} offer emulation-based solutions.
Emulation can provide detailed insights into specific software--hardware interactions but often yields large amounts of extraneous data not directly relevant to attacker--defender modeling.
By contrast, hardware and software configurations are treated as \emph{endogenous} in \emph{CyGym}, being represented through abstracted device features.
None of the above tools feature a native integration with \texttt{Gym}, complicating efforts to train and evaluate reinforcement learning (RL) agents.

Our work is closest to \cite{standen2021cyborg}, which introduced \emph{CybORG}, a Gym designed for autonomous cyber-agent development.
\emph{CybORG} combines simulation and emulation to enable rapid training, supporting multiple scenarios at varying fidelity levels.
However, it primarily focuses on training agents in simulated environments and then validating them in emulated environments, imposing specific assumptions about attackers (e.g., three hosts in two subnets, predefined software/OS access).
In contrast, \emph{CyGym} makes no such assumptions, allowing any number of devices and subnets, limited only by available computational resources.

\paragraph{Pure Simulation Environments.}
While many of the above systems rely on at least some emulation components, there exist ``pure'' simulators that do \emph{not} incorporate real operating systems.
For example, FlipIt~\cite{vanDijk2013flipit} and various \emph{Attack Graph} frameworks~\cite{durkota2015optimal,jain2011double} model adversarial actions and defenses abstractly,
focusing on strategic interactions without reproducing the full OS or network stack.
Likewise, \emph{CyberBattleSim} ~\cite{microsoft2021cyberbattlesim} offers a graph-based simulator for attacker--defender interactions using state transitions rather than emulated hosts.
These approaches can reduce extraneous data and increase scalability, but may sacrifice some realism.
\emph{CyGym} aligns more closely with these purely simulated methods while also supporting flexible device definitions and network topologies that facilitate RL research.

\subsection{Game-Theoretic Analysis in Cybersecurity}

The applications of game theory to cybersecurity have received considerable attention in the last decade~\cite{an2017stackelberg,laszka2014survey,nguyen2009security,nguyen2013analyzing,sinha2018stackelberg}.
These range from relatively simple player action sets to combinatorial actions, such as paths in a graph~\cite{durkota2015optimal,israeli2002shortest,jain2010security,jain2011double,salmeron2009worst} or plans~\cite{letchford2013optimal,panda2017near,panda2018scalable,vorobeychik2020plan}.
Reinforcement learning techniques and double-oracle methods have proved especially useful in these settings~\cite{jain2011double,nguyen2018deep,sinha2018stackelberg}, with some efforts merging these techniques into frameworks that approximate equilibria in large-scale games~\cite{bighashdel2024policy,lanctot2017unified,tong2020finding}. 
We build on this line of work, but focus on a setting that involves non-zero-sum objectives and asymmetric information, reflecting realistic attacker--defender mismatches.

Our setting also echoes the \emph{FlipIt}~\cite{vanDijk2013flipit} and \emph{FlipDyn}~\cite{banik2024flipdyngraphsresourcetakeover} games, wherein attackers aim to stealthily take control of a resource while defenders attempt to maintain or reclaim it.
However, these games remain highly stylized abstraction, whereas \emph{CyGym} provides a more comprehensive and flexible environment,
featuring arbitrary numbers of devices, subnets, and complex attacker--defender interactions that can be modeled as non-zero-sum.

In the domain of security games and learning under information constraints, two complementary strands of work are particularly relevant.  First, \cite{Sanjab} introduce a bounded‐rationality framework for two‐player zero‐sum games on cyber‐physical systems by modeling the attacker as belonging to one of three “thinking levels” (random, load‐based, or full‐optimal‐power‐flow), each of which induces a distinct decision rule.  The defender then selects a single pure strategy to maximize a weighted expectation over these attacker types, yielding a closed‐form, one‐shot equilibrium analysis.  While this approach affords clarity when attacker behavior is well‐approximated by a small set of heuristics, it requires manual specification of each “level” and produces a deterministic defense choice rather than a mixed strategy.  Second, \cite{DBLP:journals/corr/abs-1103-2491} propose a family of heterogeneous, fully distributed learning algorithms for zero‐sum stochastic games with incomplete information, in which each player follows its own reinforcement‐learning scheme (e.g., replicator, logit, or combined payoff‐and‐strategy updates).  By mapping the stochastic updates to their ODE counterparts, they prove almost‐sure convergence to saddle‐point equilibria without imposing a priori “levels” of reasoning.  In contrast to the discrete cognitive‐hierarchy approach, these heterogeneous learning schemes allow arbitrary information or computational restrictions to be encoded directly into each agent’s update rule, and yield mixed‐strategy equilibria derived via coupled stochastic approximation.  Together, these works illustrate two ends of the modeling spectrum—finite “behavioral types” versus continuous, ODE‐based learning dynamics—each of which offers insight into how bounded rationality can be incorporated into security‐game analysis.

\section{Cybersecurity Simulator}
\label{S:simulator}
%\subsection{CyGym Simulator}

At the high level, our cybersecurity simulator is comprised of four components: 1) a (hierarchically) networked set of devices, 2) workloads, 3) exploits, and 4) attack mitigation actions.
In particular, the analysis is centered around a network comprised of subnets;  each subnet, in turn, consists of a network of devices.
As such, the key unit of interest in our simulator is whether or not particular devices are compromised at particular points in time.
Some devices may be more valuable (for example, servers), and some compromises of these more severe than others (for example, exfiltration of trade secrets compared to a requirement to reboot the device).
Generally, we wish to ask questions of the form ``how many devices, on average, have been compromised between times $X$ and $Y$?'', ``what is the likelihood of particular secrets being exposed to the malicious actor?'', ``what is the expected reduction in productivity as a result of cyber-attacks?'', and the like, as we experiment with cyber-attack and defense settings.

Next, we drill down into each of the five main components: 1) the model of devices, 2) the networks and subnetworks connecting these, 3) workloads, 4) exploits (including vulnerabilities), and 5) attack mitigation actions.

\subsection{Devices}

A device is comprised of two sets of attributes which we will refer to as \emph{state}.
The first set are the attributes of the device itself, such as the type of device (e.g., server, desktop, laptop, tablet, smart phone, printer, etc), operating system and version (e.g., MacOS 12.4, Windows 11), the set of applications and versions, and any pertinent user permission information (e.g., which user accounts are on the device).
For a device $i$, we refer this set of attributes by a binary vector $d_i$.

The second set of device attributes comprise its \emph{compromise state}.
This can be as simple as a single binary attribute indicating whether the device has been compromised or not, or as complex as detailing types, levels, and history of compromise by different malicious entities.
Typically, we will capture (a) which attack has compromised the device and (b) at what access level (user or root).
We denote the compromise state of a device $i$ by a binary vector $c_i$.
The full device state for a device $i$ is then $x_i = (d_i,c_i)$.

In general, device state $x_i$ will evolve over time.
Here, we assume that time is discrete (e.g., minutes, hours, or days), with $t$ denoting a particular time step of the simulation.
Thus, we will refer to the full state of device $i$ at time (step) $t$ as $x_{it} = (d_{it},c_{it})$.
Let $X = X^d \times X^c$ be the space of possible  states, with $X^d$ and $X^c$ the space of device attribute and compromise states, respectively.
Notably, it will be important later that $d_{it}$ is observable to the defender (network administrator), but not necessarily to the attacker (unless they learn features of it through reconnaissance, say).
On the other hand, $c_{it}$ will be observable to the attacker, but not to the defender unless the latter takes concrete steps to discover it, for example, through malware and anomaly detection.

\subsection{Networks and Subnetworks}

Devices are connected via networks and subnetworks within which they are nodes.
Networks in the simulator are directed and represent which devices can access which on the network (e.g., using a valid set of credentials).
In our simulation platform, we support two ways in which such networks can be specified: exogenously (e.g., from a data file representing a real organizational network), and using a stochastic network generation model.

\subsubsection{Network Structure}

Let $D_t$ represent the set of devices in the organization at time $t$.
Additionally, we can abstractly represent devices outside the organization (that attempt to access it) as a (small) collection $\tilde{D}_t$; thus, any device in $\tilde{D}_t$ can represent a collection of these, or particular devices that the defender has specifically characterized (such as by its mac address, etc).
Recall that each device $i \in D_t \cup \tilde{D}_t$ is associated with state $x_{it}$ (although the defender only knows device configurations $d_{it}$ for devices in $D_t$).
For simplicity of exposition below, we will suppose that the set of devices $D_t$ is comprised of both those in the organization and outside.

\subsubsection{Network Dynamics}

Suppose we are given an initial network and device structure $(D_0,E_0)$, and its the hierarchical network representation $\{(V^\ell_t,E^\ell_t)\}$ that it induces using a hierarchical subnet partition model discussed above.
The next consideration is the model of network dynamics.
In particular, this entails modeling the evolution of nodes, as well as the evolution of connectivity.
In the case of the former, we assume that nodes arrive based on a stochastic generation counting process (e.g., Poisson) in discrete time, with the parameter of the process (such as the average number of new nodes arriving at each time step, or the arrival frequency) specified as a simulation configuration parameter.
An arriving node is then stochastically connected to others using a predefined network formation model (for example, a similar network connectivity model as used for stochastic network generation as discussed below).
Similarly, nodes stochastically leave, with each node associated with a probability of leaving that is, again, a simulation parameter.
In addition to the stochastic dropoff, however, nodes can be actively \emph{removed} from the network by the network administrator as part of their attack mitigation actions below (for example, a reset action can take a device offline for a period of time).
%\textbf{TODO: ensure that node removal is explicitly among the attack mitigation actions, and/or reset can have a specified duration (including inf). }

\subsubsection{Stochastic Network Generation}

The final key issue in the network model is how to obtain the initial state of the network $(D_0,E_0)$.
As with the partition, one approach that we support is to simply treat it as exogenous, for example, read from a file that describes the organizational network.
For experimental studies, however, that is insufficient, and we therefore make use of the Barabási-Albert stochastic network formation model as an alternative.

The Barabási-Albert model \cite{doi:10.1126/science.286.5439.509} is notable in that it can generate scale-free networks, which are characterized by a power-law degree distribution. This feature is crucial for modeling real-world networks, as many natural and human-made networks, including computer networks, exhibit scale-free properties. The directed Barabási-Albert graph is generated by simulating a preferential attachment process, wherein new nodes are more likely to connect to existing nodes with higher degrees. This results in a network with a few highly connected nodes (hubs) and many nodes with fewer connections, mirroring the hierarchical and uneven structure often seen in enterprise networks.

By employing the Barabási-Albert model, our framework ensures that the generated graphs are not only congruent with practical use cases but also adaptable to various network sizes and complexities. Furthermore, the directed nature of the graph supports the modeling of asymmetric relationships between nodes, which is a common characteristic in enterprise networks where data flow and access permissions are typically directional.

\subsection{Workloads}

An important feature of our simulation model is the notion of \emph{workloads}.
The goal is to enable us to capture the loss in productivity (reduction in completed workloads) that results both from cyber defense (attack mitigation) activities, as well as the attacks themselves (e.g., denial-of-service).
While productivity is typically a complex multi-dimensional consideration, we use workloads as a relatively simple abstraction which, we suggest, suffices for a game-theoretic analysis of security encounters.
%This is an important abstraction, as for game-theoretic analysis, it is likely unnecessary to consider usual productive work performed on a network at a lower level of abstraction than this.
%On the other hand, we do need a way to measure \emph{disutility} both of security mitigations that may take down the network for a period of time, as well as of successful compromise activities that disrupt network (device) functioning.
%Our model of workloads allows CySim to capture both of these features.

Specifically, a workload is a tuple $w = (X_w,\tau_w,v_w)$ where $X_w \subseteq X$ is a subset of device states $x$ within which the workload can be effectively executed (e.g., a specification of hardware, and software requirements, such as GPU types and operating system), $\tau_w$ the number of time steps the workload takes to execute, and $v_w$ the (economic) value of successfully executing the workload $w$ to the defender.

Each device $i$ is associated independently with a distribution $P_w$ over workloads, that is, the joint distribution over requirements ($X_w$), durations, and values.
We assume that each of $X_w$, $\tau_w$, and $v_w$ are independently distributed, but the duration may depend on the device state at the time that the workload was spawned.
A useful special case is when $\tau_w=1$ and $v_w = 1$, and where we restrict configuration requirements to be ranges of several device parameters only, such as OS type and version range.
These can then be generated uniformly at random.

The key feature of our workload model is the model of workload execution.
If a device $i$ generates a workload $w$ and its configuration $x_i \in X_w$, the workload successfully executes and we (the defender) accrue the associated value $v_w$.
However, suppose the device cannot support the workload (i.e., $x_i \notin X_w$)?
In this case, the device checks whether any of its out-neighbors in the network $G_t$ (i.e., any other devices that it has access to) can execute $w$.
If so, the workload is successfully executed and $v_w$ accrues to the defender; otherwise (if no out-neighbors can run it), the workload fails.
We assume that each device additionally has a workload execution limit $B_i$, so that at most $B_i$ workloads can execute simultaneously on it at any time $t$.
If a device is already running $B_i$ workloads, the effect is as if $x_i \notin X_w$ (i.e., it cannot execute the new workload).

%\textbf{We experiment with several workload distributions in Section~\ref{S:exp}.}
%\textbf{TODO: exp with different workload distributions?}

\subsection{Vulnerabilities and Exploits}

In our simulation model, we do not make a fundamental distinction between vulnerabilities and exploits.
Instead assume that each vulnerability has an exploit available for it, while also modulating the likelihood of exploit success in exploiting the vulnerability.
An attack in our model can be viewed as an effective dynamic chaining of exploits, each with a stochastic effect that may change the state of the device (for example, to compromised) according to an associated probability distribution.
Consequently, exploits constitute the core action arsenal for attackers.

We define an exploit $e$ according to the semantics similar to those for workloads (it is, after all, a unit of computation).
In particular, an exploit is a tuple $e = (X_e,\tau_e,c_e,p_e,\Delta x_e)$ with $X_e$ representing device state (e.g., versions, OS, etc that is exploitable by a particular associated vulnerability), $\tau_e$ the time that an exploit takes to execute, and $c_e$ the cost it has upon the defender.
Equivalently, we also view $c_e$ as the direct value that the exploit has to the attacker.
Finally, $p_e$ is the probability that the exploit executes successfully, and $\Delta x_e$ is its effect on state, where if state is $x$ at the time of the exploit succeeds (after $\tau_e$ steps), then the next state is $x' = x \vee \Delta x_e$, where $\vee$ is a logical OR.
The set of exploits in our simulator is populated using the NIST vulnerability database (NVD), and we take a subset of these randomly (with a predefined number of exploits) for a given simulation (although it can be configured to use all, or using alternative subsampling schemes).

From an attacker's vantage point, an exploit $e$ can be targeted at particular devices $i$ on the network.
Formally, let $a$ denote the attack actions.
We let $a_e(S)$ be an attack that attempts to execute an exploit $e$ on devices in a set $S$.
Success of the execution on a device $i \in S$ depends on 1) whether the device configuration (i.e., vulnerabilities) support execution of the exploit, that is, whether $x_i \in X_e$, and 2) an exogenous success rate $p_e$.
In particular, if $x_i \notin X_e$, the exploit fails.
If $x_i \in X_e$, then the exploit succeeds with probability $p_e$.

\subsection{Modeling Zero-Day Vulnerabilities} 

Zero-day vulnerabilities remain a fundamental challenge in security.
These are inherently difficult to defend against, as the defender (by definition) has no knowledge of them at the time that defensive actions (including mitigation against future attacks, scanning approaches, anomaly detectors, and so on) are taken.
A notable feature of the proposed simulator is that it makes modeling and reasoning about zero-day vulnerabilities quite natural, enabling us to obtain broad insights about how to defend against these.
Specifically, a key assumption in both conventional decision-theoretic and game-theoretic models of security is that both players have full knowledge of one-another's action sets, an assumption that is clearly violated by zero-days.
We address this as follows.
Suppose that $\mathcal{A}$ is the set of all attacker actions.
Let $\mathcal{A}^d \subseteq \mathcal{A}$ be the set of attack actions \emph{known to the defender}.
Thus, $\mathcal{A}^z = \mathcal{A} \setminus \mathcal{A}^d$ is the set of zero-day attacks (for example, exploiting zero-day vulnerabilities).

An additional important feature of zero-day attacks is that once they are deployed, and subsequently discovered and analyzed by the defender, they cease to be zero-day attacks and are added to $\mathcal{A}^d$.
Consequently, the nature of attack actions known to the defender is itself dynamic, and characterization of discovered attacks is a critical feature of any defense.

We discuss how to formally integrate zero-day attacks within this framework into a game-theoretic model in Section~\ref{S:game}.

\subsection{Attacker Reconnaissance}
\label{S:attackerprobe}

An important class of attacks involve reconnaissance (or \emph{probe}) actions which aim to discover information about target users, devices, and network.
Each probe $q = (X_q,p_q,J_q,o_q)$ has a set $X_q$ of configurations and probability $p_q$ of effectiveness.
A probe initiated on device $i$ succeeds on device $j$ if and only if attacker has compromised $i$ (encoded in $c_i$) and $i$ has access (directed link) to $j$ in $G_t$ at the time $t$ it is executed.
Moreover, just as with exploits, if a device $j$ being probed does not support the probe (e.g., does not respond to external pings), i.e., $x_j \notin X_q$, the probe fails, and it succeeds otherwise with probability $p_q$.

If the probe from device $i$ succeeds on a device $j$, the attacker obtains an observation (information) $o_q$ about a subset of state features $J_q$ on the device (e.g., whether a particular port is on, which OS and version is installed, etc).
This, in turn, allows a rational attacker to perform posterior inference about the device configuration $x_j$.

\subsection{Attack Mitigation Actions}

\subsubsection{Attack Detection}

%We emphasize that the defender cannot determine the compromise state with any actions, yet the defender has access to a \textit{scan} action. Our \textit{scan} action applies an Isolation Forest \cite{Liu2008IsolationF}. 
The problem of detecting attack patterns---whether it is individual pieces of software (malware) or network noise produced by malicious access patterns---is foundational in cybersecurity.
Interestingly, in simulation tools and game-theoretic analysis, this problem is commonly highly abstracted.
Commonly, for example, one posits an availability of a ``detect'' action and the like which detects an ongoing attack with some exogenously specified probability.
However, this kind of abstract analysis fails to capture critical considerations having to do with the tradeoff both the attacker makes in trading off attack aggressiveness and detectability, and the defender's own tradeoff between detection efficacy and consequences of false positives on system productivity.

To address this issue, our simulation tool incorporates machine learning based detection techniques which enable detection efficacy to be endogenous to the interaction between the defender and the attacker.
This is done by leveraging the fact that we explicitly model a workload process and distribution (see above), which provides a natural starting point for devising an \emph{anomaly detector}.
Specifically, one approach we can take is to allow a burn-in period prior to any attacks in which the defender uses workload network patterns induced by the typical network communication arising from workloads that are generated and passed among devices.
Given a numerical representation of such patterns (for example, as a time series), we can make use of any anomaly detection techniques, such as the Isolation Forest \cite{Liu2008IsolationF} (the technique we focus on in our experiments below).

%This model is trained during a burn-in time prior to attacker action. It takes the device to device communication generated from passing workloads that devices cannot accept and uses that as training data for "usual" network communication. 
When the attacker can finally take malicious actions, both their \emph{probing} and \emph{exploit} attempts generate network traffic akin to workloads, but likely distributed distinctly from these.
The attacker's decisions about which actions to take thereby impact their efficacy both in a way that is inherent to the actions (e.g., likelihood of an exploit being successful) as well as in the way they impact detectability.
Detection of attacks, in turn, provides (uncertain) information about potential attacks to the defender, which can now be used as information in follow-up defensive actions, such as to reset devices that appear to be the source of malicious traffic (with associated consequences for productivity).

In addition to detecting network anomalies, we can also target anomaly detection at particular devices.
To this end, we can similarly collect network traffic data coming into and out of the device, both during a burn-in period (when the traffic is assumed to be normal) to train the detector, and then at execution time when attacks are possible.

\subsubsection{Defensive Probing}

Since the defender does not know much a priori about the configuration of devices $\tilde{D}_t$ outside the organization, they can leverage the same reconnaissance techniques as the attacker to obtain information about these (see Section~\ref{S:attackerprobe}).

%Attacker actions \textit{probe} and \textit{attack} produce network traffic from compromised devices to other devices in the network. These communications are input to the Isolation Forest which flags a device as potentially compromised. 

%\textbf{TODO: anti-virus scans; anomaly detection; IDS}
%On detection the device is cleaned, although the defender never knows for sure if it was compromised for not, except implicitly through $u_d$. Since our framework is modular, practitioners can easily replace this with any particular scanning tool or other such detection protocol specific to their enterprise.

\subsubsection{Checkpointing and System Reset} If a system has been affected by malware, remediation inevitably involves a kind of ``reset'', which restores the system to a prior state.
Ideally, this prior state is malware-free, but of course there is always uncertainty, since detection is imperfect and the defender does not necessarily know whether (or which) malware is on the system.
%such as cleaning the malicious software from the system.
%How dramatic this process is depends on the situation, and in some cases the system may need to be rolled back in time prior to infection.
We capture this space of options through two general classes of defensive actions: checkpointing and reset either of individual devices, or the entire network (i.e., all devices).
In our model, we abstract away some of these details but aim to maintain the fundamental tradeoff between the decision to reset the system to a prior state, and the associated loss of productive activity.
Specifically, the defender can choose to execute a checkpoint action at each point in simulation time, which prevents execution of productive workloads for its (exogenously specified) duration, but saves all workloads and associated values successfully executed up to the current state.
Formally, a checkpoint action $\chi(t) = (x_t,\delta)$ executed at time $t$ is a tuple of the device state $x_t$ preserved by it as well as the duration, so that no workload can be executed between time $t$ and $t+\delta$.
We let $t_\chi$ denote the time of a checkpoint $\chi$ and similarly use $x_\chi$ to denote the device state saved in checkpoint $\chi$.
Let the set of available checkpoints in time step $t$ be $C(t)$.
We additionally assume that $C(t)$ has size at most $K$, and if we add a checkpoint that would cause $C(t)$ to exceed $K$, the oldest checkpoint in $C(t)$ is removed.

A reset action, in turn, which also prevents workload execution for a prespecified duration, rolls the system back to a specified checkpoint.
Formally, this is $\rho(t) = (\chi,\tau)$, where $\chi \in C(t)$ is a checkpoint to which the system reverts, while new workloads cannot execute between $t$ and $t+\tau$.
The system state after $\rho(t)$ is $x_\chi$.
Moreover, the reset action $\rho(t)$ removes all workloads that have been generated on this device and successfully executed between time $t_\chi$ and $t$.

%\subsubsection{Software Updates} 

\subsubsection{Device Reconfiguration and Software Update} 

Another tool in the defender's toolbox involves device-level reconfiguration, which may entail removing some of the software installed, adding software (e.g., anti-virus that would prevent certain malware from being run), blocking ports, etc.
Let $D_r \subseteq X^d$ be the set of possible reconfigurations to the device state $d_i$ that can be implemented on a device at any particular point in time.
Note that the impact of reconfiguring a device is reflected in productivity endogenously, since some of the spawned workloads will require particular configurations to execute.

An important decision within the broad class of device reconfigurations that a network administrator needs to make involves the nature and frequency of software updates.
Any update is at least somewhat disruptive to productivity.
Moreover, updates may themselves introduce new vulnerabilities, while at the same time patching old ones.
We capture this tradeoff by explicitly modeling software updates as defender actions.
Specifically, let $u = (X_u,\tau_u,\Delta x_u)$ be the tuple representing a software update, where $X_u$ is the set of configurations (e.g., software, OS) in which the update can be installed, $\tau_u$ the duration of installation (so that no workload can be spawned/executed between time $t$ when the update begins and time $t+\tau_u$), and $\Delta x_u$ is the state change resulting from the update, i.e., if the device state is $x$, after the update it becomes $x' = x \oplus \Delta x_u$ where the $\oplus$ (logical XOR) operator represents the change in state (implemented as flipping the bits in state corresponding to the old and new version of the software, OS, etc, for example).
Note that in practice, $\Delta x_u$ will only impact the device state $d_i$, and not its compromise state $c_i$.

\subsubsection{Network Reconfiguration} 

A final set of defensive tools involves reconfiguring the network.
We consider two kinds of network changes: device removal (actually physically removing the device, or blocking its access to the network, black listing, etc) and edge removal (blocking access between devices).
To formalize, at any point in time, the defender can select a subset $S \subseteq D_t \cup \tilde{D}_t$ devices to remove from the network (e.g., black list).
In the case of the latter, we similarly allow the defender to remove a subset $L \subseteq E_t$ of edges from the network.

%TODO (block access to some devices from others -- firewall; white lists vs black lists)

\section{Simulation-Based Game-Theoretic Model}
\label{S:game}

We now use the simulation model discussed above to formally define a \emph{simulation-based game-theoretic model} of cybersecurity encounters.
First, there are two players: the defender $\beta$ (e.g., network designer, administrator, etc) and the attacker $\alpha$; these are already explicitly referenced in the simulator described in Section~\ref{S:simulator}).
Second, the game takes place over a sequence of discrete time steps $t$ starting at $t=0$ and ending at a predefined horizon $T$.
At time $0$, the attacker has no knowledge of the network structure $G_0$ or device configurations $d_i$, and we assume some initial proportion $\eta$ are compromised, all others are clean (i.e., compromise state $c_{i0}$ is a zero vector for all other devices, 1 for the fraction $\eta$).
Additionally, we assume that the compromised state $c_{it}$ at any time $t$ after a successful compromise is known to the attacker but not the defender.
However, we assume that there is a common knowledge distribution (e.g., stochastic model) of the network generation and device configuration, i.e., $G_0 \sim \mathcal{D}_G$ and $\{d_i\}_{i \in D_0} \sim \mathcal{D}_d$.
Consequently, our game entails \emph{partial observability} of game (device) state by both players.

%Additionally, the simulation model explicitly describes actions available to each player.

The attacker has two types of underlying actions they can take at any time step, the semantics of which are described in Section~\ref{S:simulator}: a) probes $P$ and b) exploits $E$.
The defender, in turn, has six types of actions: a) detector execution, b) defensive probing (of external devices), c) checkpointing, d) reset, e) device reconfiguration (including software updates), and f) network reconfiguration.
All but probing actions, when they succeed (which can be stochastic), have a direct impact on the system state in the next time step as detailed below and in Section~\ref{S:simulator}; an exception are the probing actions by both players, which impact solely the information available to each player about the state, but do not change the system state directly.
Both the defender and the attacker can select at most one of these actions at any point in time, but may also do nothing, which corresponds to a null action that has no consequences $a^\alpha$ and $a^\beta$ for the attacker and defender, respectively.

At each time step, both players choose actions simultaneously, but these are executed (for the purposes of computing the next state and rewards) by implementing first the attacker's action followed by the defender's action.
Both the dynamic nature of the game and imperfect observability of state by both players make our model an instance of a \emph{partially observable stochastic game (POSG)}.

We describe the details of the attacker and defender actions in each category for the attacker and defender next, and subsequently discuss our model of defender and attacker rewards in the POSG.

\subsection{Attacker Actions}

\paragraph{Attacker Probe Actions}

Each attacker probe action $a_{q,i,j}^\alpha$ is associated with a particular probe $q$ which targets device $j$ from device $i$ (i.e., there is a distinct action for each probe $q$ and device pair $i$ and $j$, although of course only a subset of these have a non-zero probability of success, and it is straightforward to prune this set accordingly to only consider probes across existing network edges).
Let $A_\mathit{probe}^\alpha$ be the set of all such (feasible) combinations (note that a probe may also be guaranteed to fail due to configuration mismatch on the target device $j$, but this information may not be available to the attacker).
The strategic decision corresponds to selecting a problem $a^\alpha \in A_\mathit{probe}^\alpha$ at a given time step $t$.
The side-effect of each probe is purely informational, providing the attacker with some of the details about the  states of devices on the network (but not complete details).

\paragraph{Attacker Exploits}

Each attacker can also execute an exploit targeted at a device $i$.
The details of exploits and their effects are described in Section~\ref{S:simulator}; the net effect of an exploit, if successful (which is determined stochastically) is to change the compromise state $c_{it}$ of the targeted device $i$ at the time $t$ the exploit is executed.
Moreover, this compromise state remains unchanged unless either the defender's actions or another exploit by the attacker directly changes it.

\subsection{Defender Actions}

\paragraph{Attack Detection} There are two kinds of actions which stem from the use of device and network anomaly detection: 1) when (at which time step $t$) to deploy them (with these remaining active thereafter), and 2) how to configure them at each time step $t$ post deployment.
In the case of the former, we assume that both a network anomaly detector, as well as device-level detectors, are deployed after an initial burn-in period of time, with the timing of deployment $t$ chosen strategically by the defender.
The tradeoff faced thereby is to delay deployment time to facilitate more collected data, while avoiding capturing malicious data that arises from attacks prior to deployment (which effective poison the detector).
In terms of configuration, we consider (a) retraining (i.e., collecting additional data post deployment which can be used to retrain the detector), and (b) modifying sensitivity (i.e., the tradeoff parameter that determines the false positive and false negative rate).
At a given time step $t$, we allow only one of these actions to be taken by the defender (deploy, modify sensitivity, or retrain and redeploy---with the latter also choosing sensitivity parameter).

As long as a detector is currently deployed (i.e., deployment time was prior to current time step $t$), we assume that its effects are fully autonomous.
This is implemented by the detectors performing their testing in each time step $t$ post deployment, and resulting alerts becoming part of available information for the defender to act on.

\paragraph{Defensive Probing Actions} Just as in the case of attacker, the defender's probing actions target particular devices that are not in the defender's network to obtain their characteristics.
Structurally, these mirror the attacker's probing actions described above, with the same constraints imposed; thus, each probe targets a device $j$ from a device $i$, so that the full set of probes corresponds to the cross product between probe action types (ping, traceroute, etc) and pairs of devices.
We let $A_\mathit{probe}^\beta$ be the set of feasible probes (i.e., requiring connectivity from $i$ to $j$).

\paragraph{Checkpointing Actions} Checkpointing actions involve saving a device state at time step $t$, as described in Section~\ref{S:simulator} above.
For the duration of this action, no workloads can be executed on the device (we can model this by having a device state variable in $d_i$ which is 1 whenever workloads cannot execute, and is then reset to 0 once checkpointing time is over).
However, checkpointing a device does not prevent the defender from executing other actions affecting this device during that time period.
The corresponding action set is then isomorphic with the set of all devices $D_t$ on the defender's network.

\paragraph{Reset Actions} Similar to checkpointing, we can execute one of a set of reset action types (described in Section~\ref{S:simulator}) on any of the devices on the defender's network.
The cross product between reset types and the device set then comprises the set of reset actions that the defender can take at any time point $t$.

\paragraph{Device Reconfiguration Actions} The device can be reconfigured at any point in time $t$ in several ways (including software updates, patching, etc) detailed in Section~\ref{S:simulator}.
Each reconfiguration type can be performed on each device, so that the set of actions in the associated game is the cross product between these.

\paragraph{Network Reconfiguration Actions} The final category of defensive actions involve network reconfigurations, which are comprised of removing directed edges and removing devices from the network.
Each action is thus a choice of either removing an edge $(i,j) \in E_t$ or removing a device $i \in D_t$, and the union of these constitutes the set of all network reconfiguration actions that the defender has in the game.

\subsection{Attacker and Defender Rewards} The next key element of the game model involves the rewards that accrue to both players in each time step $t$ of the game.
We begin with the attacker.
We associate with each attacker action $a^\alpha$ a cost $\zeta^\alpha(a^\alpha)$.
%For example, zero-day exploits entail development cost which is likely much higher than the expected cost of executing one of the known exploits, while the cost of probes relatively small compared to any exploits.
For each device $i$, the attacker further receives a reward $r(x_i) \ge 0$ (additive over devices) which is a function of device state $x_i$; we assume that $r(x_i)=0$ if the device $i$ has not been compromised by the attacker, and otherwise depends on the nature of the configuration (e.g., whether it contains sensitive data, has crashed, executes malicious application, etc) and compromise (e.g., user-level or root).
Probes have a positive reward when discovering new devices. However, this creates a traffic pattern that the defender can exploit to determine lateral movement as descibed in section 3.7.
The instantaneous net utility of the attacker is then
\[
r_A(a^\alpha,x) = \sum_i r(x_i) + \zeta^\alpha(a^\alpha),
\]
where $x$ is the true state of all devices on the network.

In the dynamic setting defined by the simulator, the attacker takes attack actions over time according to a \emph{policy}, which in general maps an attacker's \emph{belief state} (distribution over the true state of all devices $x$ to actions $a^\alpha$ in each time step of the game.
In practice, fully dealing with and update such a belief state is intractable, and we instead make use of (a finite history of) observations, such as compromise state and any information about devices obtained through probes.
We denote this observation history by $o^\alpha$.
The attacker's policy $\pi^\alpha$ then maps $o^\alpha$ to an action.

For the defender, each action is associated with a non-negative execution cost. %(similar to the attacker).
Rewards, on the other hand, stem from executing workloads, with each successfully executed workload $w$ accruing a reward of $r = v_w$ to the defender.
%If a reset action rolls back a device $i$ to a prior state, all accrued rewards spawned on this device (even if executed elsewhere) are also rolled back (which we can equivalently implement as a negative reward that subtracts all the associated values $v_w$).
If a reset action rolls back a device $i$ to a prior state, all current workloads $w$ are dropped and device $i$ becomes busy for $t \sim \mathrm{Triangular}(m_{\textit{min}},m_{\textit{mode}},m_{\textit{max}})$ time steps with $m_{\textit{min}},m_{\textit{mode}},m_{\textit{max}}$ being game parameters.
This means that the defender's utility depends on the full trajectory, rather than each step independently.
In addition, the defender loses the amount $r(x_i)$ that the attacker gains for each device.
As in the case of the attacker, the defender's policy $\pi^\beta$ will map subjective observations $o^\beta$ to actions $a^\beta$.

Given a policy pair of the defender and attacker, the expected utility of the attacker over a finite horizon $T$ is
\[
U^\alpha(\pi^\alpha,\pi^\beta) = \mathbb{E}\left[\sum_t r_A(\pi(o_t^\alpha,x_t))\right],
\]
where the expectation is with respect to any uncertainty in the system, including that induced by the information available to the attacker, as well as the joint player policies.
Similarly, the expected utility of the defender is
\[
U^\beta(\pi^\alpha,\pi^\beta) = \mathbb{E}\left[R(\tau) - \sum_t \zeta^\beta(\pi^\beta(o_t^\beta))\right],
\]
where $\zeta^\beta$ is the defender's cost of executing an action, $\tau$ is a trajectory which comprises the set of workloads executed, denoted by $W(\tau)$ along with a sequence of device states $x_t$, and $R(\tau)$ the reward of the trajectory defined as
\[
R(\tau) = \sum_{w \in W(\tau)} v_w - \sum_t \sum_i r(x_{i,t}).
\]

\subsection{Modeling Zero-Day Exploits} 

As we discussed in Section~\ref{S:simulator}, one of the core challenges in cybersecurity modeling is how to credibly capture zero-day attacks.
In the realm of game theory, this issue is particularly acute, as fundamental to conventional game modeling is the assumption that the \emph{strategy space} of the different actors is known and given.
Of course, if we resolve strategies of attackers at sufficiently fine granularity (for example, software design, experimentation, and so on), we can in principle capture zero-day development as well, but this is typically too complex to be helpful.
In the simulator itself, we described that we can model zero-days by subsampling the full space of exploits to only consider a subset of these as part of the active strategic space of the attacker---as far as the defender knows, that is---with others effectively unknown, but strategically deployable by the attacker as they wish (at which point they become discovered).
The challenge now, however, is how we can formally incorporate this structure into the game model.

To do this, we propose that there is a commonly known distribution over the space of \emph{all possible exploits} as defined by exploit parameters in Section~\ref{S:simulator} (for example, a uniform prior at time step $t=0$ of the game).
Formally, let $z$ be a parametric representation of exploits; for example, it can be a binary encoding of OS, application, version range, and vulnerabilities that are exploited.
Let $D_z$ denote the distribution over $z$, which we assume is known to both players.\footnote{This requirement of a common knowledge prior is central to a formal game-theoretic model, but is also not a strong requirement, since we can assume it to be an uninformed (uniform) prior.}
The game thus begins with a known set of exploits (to both the defender and the attacker), and this common knowledge distribution over exploits (attack actions) that the attacker may or may not have.
In general, since we may not know \emph{how many} zero-day exploits the attacker actually possesses, we may additionally have a prior distribution over the number $N_e$ of these.
To simplify our discussion, suppose $N_e = 1$---that is, an attacker has a single zero-day, and only its parameters $z$ are unknown.

As soon as a zero-day exploit is executed, we assume that the defender's posterior over it becomes 1, which in practice simply means that it is added to the defender's knowledge of which actions are available to the attacker, thereby potentially substantively changing the strategic behavior of both players.
Moreover, the history of zero-days may influence the defender's posterior distribution over the characteristics of these (which may, in turn, create an opportunity for deception on the part of the attacker in strategically choosing which to execute in which order).
From a game-theoretic perspective, this is well-defined, but induces considerable strategic complexity insofar as the attacker's strategy space becomes a function of available sets of exploits over which the defender has only distributional information.

To formalize the information asymmetry inherent in zero-day attacks, we let the attacker policy be indexed by $z$, that is, we denote it by $\pi^\alpha(o,z)$, whereas the defender's policy remains as before, since $z$ is unknown to the defender.
Let $\mathcal{E}$ be the set of exploits the attacker has, not including the zero day, and let $\mathcal{Z}$ be the set of all \emph{possible} zero-day exploits.
Further, let $e(z)$ denote the exploit with parameters $z \in \mathcal{Z}$.
Suppose that the actual zero-day exploit is parameterized by $z$.
Then the attacker cost of executing it is $\zeta^\alpha(e(z);z)$, while the cost of executing any $z' \in \mathcal{Z}\setminus z$ (i.e., any exploit it does not in fact possess) is $\zeta^\alpha(e(z');z) = \infty$.
The utility model of the defender above assumed that the set of exploits is known and fixed.
We can make that explicit by denoting it by $U^\beta(\pi^\alpha,\pi^\beta;z)$, and the expected \emph{ex ante} defender utility is then $\mathbb{E}_z[ U^\beta(\pi^\alpha,\pi^\beta;z)]$, where the expectation is with respect to $D_z$.
Since the attacker knows its zero-day, the attacker's utility can be explicitly a function of $z$, i.e., $U^\alpha(\pi^\alpha,\pi^\beta;z)$, defined as above.

\subsection{Bayes-Nash Equilibrium of the POSG}

In the POSG, the underlying player strategies are the policies.
Let us restrict consideration to the policy sets $\Pi^\alpha$ and $\Pi^\beta$ for the attacker and defender respectively, mapping observations to actions as discussed above.
Further, let $\mathcal{P}(\Pi)$ denote the set of all probability distributions over a set of policies $\Pi$, and let $\sigma \in \mathcal{P}(\Pi)$ denote a probability distribution (i.e., a \emph{mixed strategy}) in this set.
The utility of each player for a pair of mixed strategies $\sigma^\alpha \in \mathcal{P}(\Pi^\alpha)$ and $\sigma^\beta \in \mathcal{P}(\Pi^\beta)$ is just the expectation of each player's utility $U^\alpha$ and $U^\beta$, respectively, with respect to the associated probability distributions:
\[
U^\alpha(\sigma^\alpha,\sigma^\beta) = \mathbb{E}_{\pi^\alpha \sim \sigma^\alpha, \pi^\beta \sim \sigma^\beta,z \sim D_z} [U^\alpha(\pi^\alpha,\pi^\beta;z)] \quad \mathrm{and}\quad U^\beta(\sigma^\alpha,\sigma^\beta;z) = \mathbb{E}_{\pi^\alpha \sim \sigma^\alpha, \pi^\beta \sim \sigma^\beta} [U^\beta(\pi^\alpha,\pi^\beta)]. 
\]

The $\epsilon$-\emph{Bayes-Nash equilibrium (BNE)} of the resulting game is a mixed-strategy pair $(\sigma^\alpha,\sigma^\beta)$ such that for each player $i \in \{\alpha,\beta\}$, 
\[
U^i(\sigma^i,\sigma^{-i}) \ge U^i(\pi^i,\sigma^{-i}) - \epsilon
\]
for all $\pi^i \in \Pi^i$.

Our goal will be to compute an $\epsilon$-BNE for a small $\epsilon$.
We describe a computational approach for this based on the well-known double-oracle~\cite{jain2011double} and PSRO~\cite{bighashdel2024policy} frameworks next.

\section{DOAR: Double Oracle with Actor Response Ascent}

We now turn to the problem of finding equilibrium behavior in our partially observed, stochastic cyber–defense game. 
A key challenge is that the strategy spaces of both players are policies, in which even the underlying action spaces (choices in each time step) are combinatorial (for example, considering all subsets of devices on the network).
A naive solution approach is to use policy-space response oracle (PSRO)~\cite{bighashdel2024policy}, a generalization of a double-oracle approach, in which we start with an arbitrary small sets of policies for both players (e.g., heuristic, random, etc), and iterate between computing a Nash equilibrium of the current game matrix, and computing best responses to the equilibrium found in the previous iteration.
In PSRO, best responses are approximated using reinforcement learning (RL).

In our setting, however, applying RL directly to compute (approximate) best responses of the players is challenging due to the combinatorial nature of actions.
%Our DOAR framework interleaves four steps each round: seed a restricted subgame with trivial (“pass”) or random policies; generate best responses to the current mixture via critic‐guided coordinate‐ascent beam search over device subsets, exploits, and scan targets; solve the restricted bi‐matrix for a Nash equilibrium mixture; and stop when no reply improves the equilibrium payoff by more than a tolerance or when a maximum number of rounds is reached (with “far-apart” restarts if either learner stalls twice). 
The major innovation in the proposed DOAR is in the way we construct best responses to deal with this combinatorial explosion issue.
In particular, we employ a \emph{critic-guided coordinate-ascent beam search} that leverages our learned Q-function to efficiently navigate the combinatorial action space (Algorithm~\ref{alg:beam}), which we now describe.  
\begin{figure}[H]
  \centering
  \scalebox{0.9}{% adjust this factor
    \begin{minipage}{1.11\linewidth}% adjust width so caption doesn’t overflow
      \begin{algorithm}[H]
        \scriptsize        % or \footnotesize, \tiny, etc.
        \caption{Critic‐Guided Coordinate‐Ascent Beam Search for Best Response}
        \label{alg:beam}
        \begin{algorithmic}[1]
          \Require current state $s$, critic network $Q_\phi(s,a)$, beam width $K$, temperature $\tau$, 
          \quad\quad number of devices $D$, action dimensions $T,E,P$
          \State Define the no–op action 
          \[
            \mathrm{noop} \;=\;(t_{\mathrm{noop}},\;\emptyset,\;\emptyset,\;0)
          \]
          \State Compute its base value $Q_{\mathrm{base}}\gets Q_\phi(s,\mathrm{noop})$
          \For{$d=1$ \textbf{to} $D$}
            \State Form per–device candidate set
            \[
              C_d \;=\;\{\mathrm{noop}\}\;\cup\;\{(t,\{d\},e,p)\mid
                t=0,\dots,T-1,\;e=0,\dots,E-1,\;p=0,\dots,P-1\}.
            \]
            \State Evaluate each $a\in C_d$: 
            \[
              Q(a)\;\gets\;Q_\phi(s,a).
            \]
            \State Keep the top-$K$ by value: 
            \[
              B_d \;\gets\;\arg\!\mathrm{top}\!-\!K_{a\in C_d}Q(a).
            \]
            \State Sample one action
            \[
              a_d\sim\mathrm{Categorical}\bigl(\exp\{Q(a)/\tau\}_{a\in B_d}\bigr).
            \]
          \EndFor
          \State Merge individual picks $\{a_d\}$ into a joint action
        \[
          D^{*} = \{\, d \mid a_{d} \neq \mathrm{noop} \,\}, \quad
          t^{*} \sim \{\, t_{d} \mid d \in D^{*} \,\}, \quad
          e^{*} = \bigcup_{d} e_{d}, \quad
          p^{*} \sim \{\, p_{d} \mid d \in D^{*} \,\}.
        \]
          \State \Return best‐response $(t^*,D^*,e^*,p^*)$
        \end{algorithmic}
      \end{algorithm}
    \end{minipage}
  }
\end{figure}
The issue that our approach addresses is that in standard actor-critic methods the actor network produces a continuous vector $\pi_\theta(s)\in[-1,1]^A$ which is then discretized by taking an $\arg\max$ over each one‐hot block\cite{xiong2018parametrizeddeepqnetworkslearning}.  In a combinatorial action space this block‐wise $\arg\max$ rarely recovers high‐value joint actions: individual bits never “compare notes,” gradients are spread across thousands of possible device–exploit combinations, and the resulting deterministic decode easily becomes trapped in local modes\cite{lillicrap2019continuous}.  In contrast, our critic‐guided coordinate‐ascent beam search uses the critic $Q_\phi(s,a)$ directly to guide a structured search: for each device $d$ we enumerate all single‐device moves $(t,\{d\},e,p)$, score them with $Q_\phi(s,a)$, keep the top‐$K$ candidates, sample one via $\operatorname{Softmax}(Q/\tau)$, and then merge those per‐device picks into a coherent joint action $(t^*,D^*,e^*,p^*)$.  By explicitly comparing and recombining high‐value pieces under the critic’s global value estimates, beam search recovers far stronger best‐responses in huge discrete spaces than naïvely decoding the actor’s continuous output. It should be noted that this works in the context of DOAR because we simply require a \textit{better} response, even if this response is not \textit{best}. That said, this per-device greedy merge can miss synergistic, multi-dimensional actions: for instance, two devices might each be only mildly vulnerable on their own, but targeting them together (or pairing a specific exploit with a specific device) could unlock a super-additive payoff that a purely independent per-device pick overlooks.

\begin{figure}[h!]
  \centering
  % Scale the entire 2‐panel block to 0.8×\textwidth (adjust as needed)
  \resizebox{0.8\textwidth}{!}{%
    \begin{minipage}{\textwidth}
      \centering
      %── first subfigure ──────────────────────────────────────────
      \begin{subfigure}[t]{0.45\textwidth}
        \vspace{0pt}% ensure top alignment
        \includegraphics[width=\linewidth]{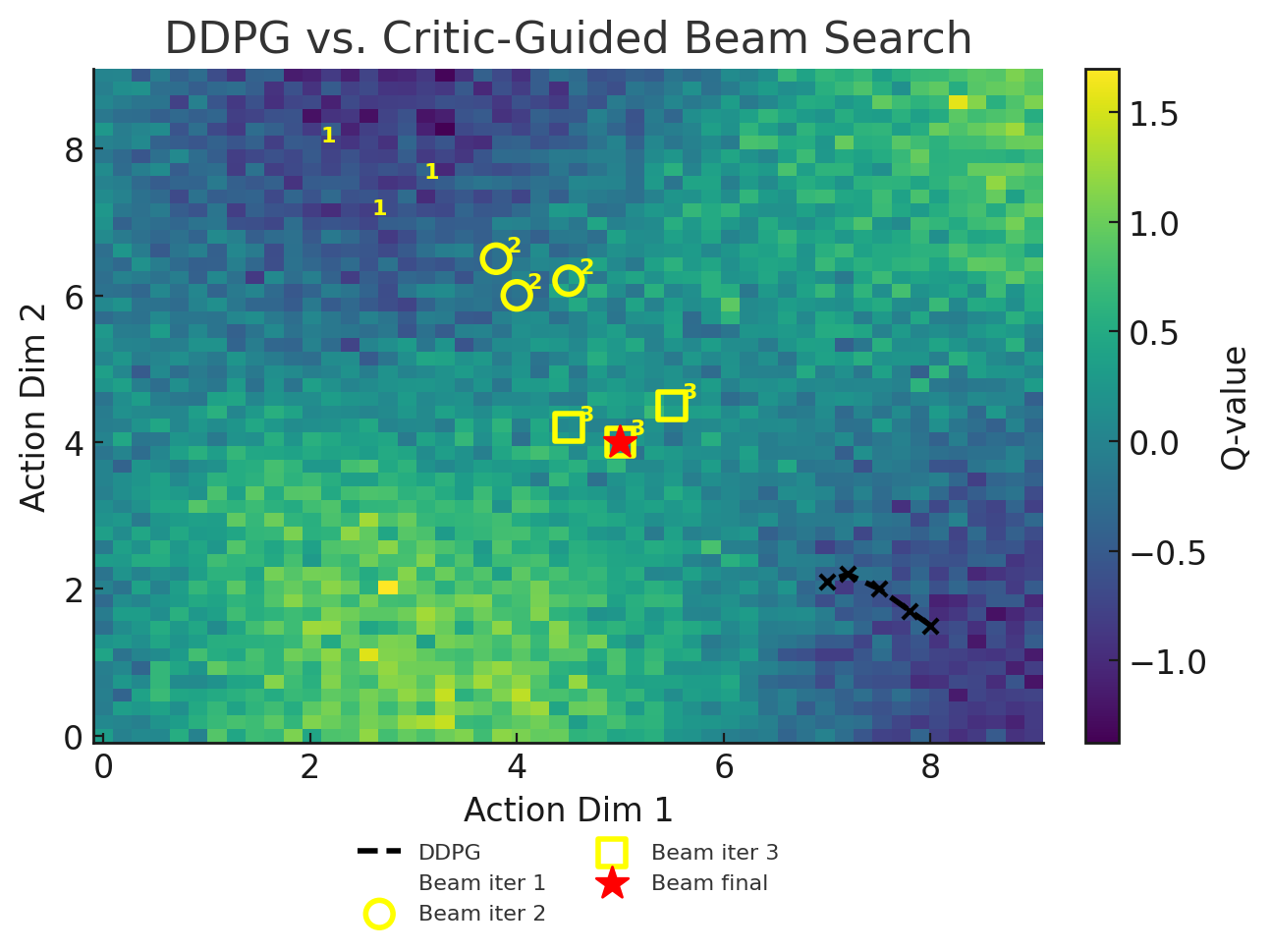}
        \caption{Comparison of DDPG (black line) versus critic‐guided beam search on a noisy 2D Q‐surface. Beam‐search candidates are yellow; final choice is a red star.}
        \label{fig:beam-vs-ddpg}
      \end{subfigure}\hfill
      %── second subfigure ─────────────────────────────────────────
      \begin{subfigure}[t]{0.55\textwidth}
        \vspace{0pt}% ensure top alignment
        \includegraphics[width=\linewidth]{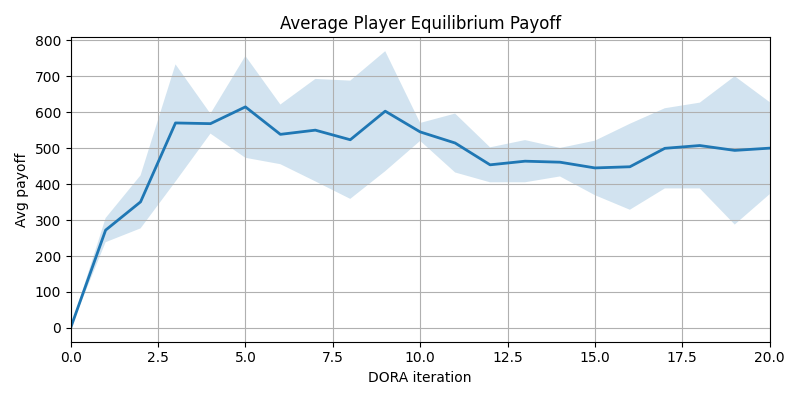}
        \caption{DOAR average equilibrium payoff over iterations (±1 std dev).}
        \label{fig:doar-payoff}
      \end{subfigure}
    \end{minipage}%
  }
  %── combined caption ─────────────────────────────────────────
  \caption{%
    (a) Beam search vs.\ DDPG on a noisy 2D Q‐surface.\quad
    (b) DOAR average equilibrium payoff over iterations.%
  }
  \label{fig:combined}
\end{figure}

As shown in Figure~\ref{fig:beam-vs-ddpg}, the standard DDPG policy (black path) quickly becomes trapped in a local region of low Q‐value, whereas our critic‐guided beam search is able to explore multiple candidate actions per device and ultimately recover a better payoff.

\section{Experiments}
\label{S:exp}

\iffalse
\subsection{Case Study (To DO?)}
%\textbf{TODO: stochastic network partitioning models (a) random, (b) depends on device configurations $d_i$ and connectivity.}

\textbf{TODO: evaluate impact of different levels of action abstraction on solutions and results. Especially, evaluate the value of abstraction in the case of detection.}
\fi

We demonstrate \texttt{CyGym} using a case study of Volt Typhoon.

%\subsection{Experiment Setup}

\paragraph{Volt Typhoon Environment}
The \texttt{Volt\_Typhoon\_CyberDefenseEnv} builds on our base \texttt{CyberDefenseEnv} to emulate the distinctive operational and threat‐model characteristics of the Volt Typhoon scenario. At initialization, the network comprises a heterogeneous fleet of Windows Server hosts—three of which are designated as domain controllers—with remotely accessible services (VPN, RDP, Active Directory, administrative password management and FortiOS) instrumented to reflect real‐world configurations. Two critical CVEs (ED3A999C-9184-4D27-A62E-3D8A3F0D4F27 and 0A5713AE-B7C5-4599-8E4F-9C235E73E5F6) are seeded across these services to model plausible exploit pathways. To capture pre-existing adversary footholds, 40\% of active hosts are randomly marked as compromised on reset, and an additional five devices are assigned “attacker‐owned’’ status. Each host executes both benign and adversarial workloads whose durations follow a triangular distribution, enabling quantitative assessment of defender work‐completion utility under attack. Defender agents may execute fine‐grained actions, each inducing explicit costs or device busy times; attacker agents may probe undiscovered hosts or launch vulnerability exploits, with compromised domain controllers guaranteed to succeed in all subsequent exploits. This design achieves a balance between scenario fidelity and tractability for reinforcement‐learning and game‐theoretic strategy synthesis.

\paragraph{State Space}

The environment tracks up to \(N_{\max}\) devices, each described by six features—OS code, version number, compromise flag, anomaly score, attacker-knowledge flag, and presence indicator—concatenated (and \(-1\)-padded) into a full state vector \(x\in\mathbb{R}^{6N_{\max}}\). The defender’s observation \(o_{d}\in\mathbb{R}^{6N_{\max}}\) masks comprimise information as well as attacker owned and devices not yet added to the network via its evolution procedure. The attacker’s observation \(o^\alpha\in\mathbb{R}^{4N_{\max}}\) further hides anomaly scores and presence indicators of the defender, drops the attacker-knowledge feature entirely, and masks inactive or unknown devices. 
The defender's observations $o^\beta$ include anomaly scores and device configurations, but not actual compromise flags.
%Hence \(\dim(o^{\beta})=6N_{\max}\) and \(\dim(o^{\alpha})=4N_{\max}\), and both agents act under partial observability.

\paragraph{Actions and Utilities}

We instantiate the actions of both players as outlined in Table~\ref{tab:combined_effects_utils}.
We assume that each workload has unit value, i.e., $v_w = 1$ for all $w$, and the reward to the attacker (and cost to the defender) for a successfully compromised device at each time step is $r(x_i)=1$.

\iffalse
Formally, at each decision step the agent selects
\[
a = (t, D, e, p)\;\in\;
\begin{aligned}
  &\{0,\dots,T-1\}
    \times\bigl(\mathcal P(\{1,\dots,N_{\max}\})\setminus\{\emptyset\}\bigr)\\
  &\quad\times\{0,\dots,E-1\}\times\{0,\dots,P-1\}
\end{aligned}
\]
where
\begin{itemize}
  \item \(t\) is the action type (there are \(T\) types),
  \item \(D\subseteq\{1,\dots,N_{\max}\}\) is a non‐empty subset of the \(N_{\max}\) devices,
  \item \(e\in\{0,\dots,E-1\}\) is the chosen exploit index,
  \item \(p\in\{0,\dots,P-1\}\) is the chosen application index.
\end{itemize}
Hence the total discrete action‐space size is \(\lvert\mathcal A\rvert = TPE(2^{N_{\max}}-1).\)
\fi

\begin{table}[h!]
  \centering
  \begin{tabular}{|l|l|p{6cm}|p{4cm}|}
    \hline
    \textbf{Role}   & \textbf{Action}     & \textbf{Effect} & \textbf{Utility (per device)} \\ 
    \hline
    \multirow{6}{*}{\emph{Defender}} 
     & Clean               & Clears compromise flag, incurs device busy‐time 
                          & $+0.30$ if compromised, else $-0.01$ \\ \cline{2-4}
     & Checkpoint          & Saves state (of all devices) to disk for future rollback 
                          & $-0.50$ \\ \cline{2-4}
     & Restore             & Restores network from last checkpoint (losing intermediate work) 
                          & $-1.00$ \\ \cline{2-4}
     & Upgrade             & Increments application version, incurs busy‐time 
                          & $-1.00$ (device also becomes busy) \\ \cline{2-4}
     & Scan                & Executes ML‐based anomaly detection on recent logs 
                          & $-0.50$ \\ \cline{2-4}
     & Block / Unblock     & Toggles network edges to control reachability 
                          & $-0.50$ \\ \cline{2-4}
     & Pass                & No action taken 
                          & $0$ \\ 
    \hline
    \multirow{3}{*}{\emph{Attacker}} 
     & Attack              & Deterministically exploits any matching vulnerability on each targeted host 
                          & $+1.00$ per compromise; $+10.0$ if via domain controller \\ \cline{2-4}
     & Probe               & Discovers previously unknown neighbors via network scanning 
                          & $+0.10$ per successful discovery \\ \cline{2-4}
     & Pass                & No action taken 
                          & $0$ \\ 
    \hline
  \end{tabular}
   \caption{Action Types, Effects and Immediate Utilities for Defender and Attacker. Utility values are subject to network‐hyperparameter scaling (See Section 6 for details).}
   \label{tab:combined_effects_utils}
\end{table}

%In brief, defender actions carry negative immediate utilities (i.e.\ costs or work‐delays), while successful attacker actions yield positive utilities (i.e.\ rewards for new compromises).  The complete per‐device utility schedule is given in Appendix A.

\iffalse
After each multi‐device action at time \(t\), utilities are computed as
\begin{align}
U^{A}_{t}
&= \bigl|\mathcal{D}^{\mathrm{new}}_{t}\bigr|
  - \sum_{j\in\mathcal{J}_{t}} c^{A}(a_{j})
  + \beta_{A}\bigl(\gamma\,\phi^{A}_{t} - \phi^{A}_{t-1}\bigr),
  \\[6pt]
U^{D}_{t}
&= \alpha\,W_{t}
  - \sum_{i\in\mathcal{I}_{t}} c^{D}(a_{i}).
\end{align}
where:
\begin{itemize}
  \item \(\mathcal{D}^{\rm new}_{t}\) is the set of newly compromised hosts at step \(t\).
  \item \(W_{t}\) is the number of defender workloads completed.
  \item \(\mathcal{J}_{t}\), \(\mathcal{I}_{t}\) index attacker/defender devices receiving actions \(a_{j}\), \(a_{i}\), respectively.
  \item \(c^{A}(\cdot)\), \(c^{D}(\cdot)\) are the per‐device immediate costs (negative utilities) for each action (see Appendix A).
  \item \(\phi^{A}_{t}=n_{t}/N\) is the attacker potential, where \(n_{t}\) is the current number of compromised hosts and \(N\) is the total network size.  \(\gamma\) is the discount factor, and \(\beta_{A}\ll1\) is a small shaping weight.
\end{itemize}
\fi

To help with training,
we additionally include the potential‐based shaping term in the attacker's instantaneous reward at time step $t$
\[
\beta_{A}\bigl(\gamma\,\phi^{A}_{t}-\phi^{A}_{t-1}\bigr),
\]
where \(\phi^{A}_{t}\) is the fraction of compromised hosts at time $t$.
%because the fraction of compromised hosts \(\phi^{A}_{t}\) is a 
%Since \(\phi^{A}_{t}\) is a \emph{monotonic} global statistic of the network state, it provides for an effective dense performance feedback to the attacker.
%This aids attacker behavior learning as we will discuss next. 
By shaping on using this one‐step change, each local attack action immediately yields feedback about its impact on the overall compromise level—information that would otherwise only arrive much later via sparse compromise‐based utilities.  This dense, global signal does not affect the long‐run equilibrium, since such potential‐based shaping is provably \emph{policy‐invariant} \cite{Ng1999PolicyIU}.  
The weight \(\beta_{A}\) is chosen small enough so that it only accelerates strategy generation.

\paragraph{Network Structure and Dynamics} The network is initialized as a connected Barabási–Albert graph seeded with domain controllers and Windows Server hosts. Initially, a fixed proportion of active nodes are marked compromised at random and a fixed count of nodes are designated \textit{attacker‐owned}; \textit{attacker‐owned} nodes remain permanently compromised and represent devices outside of the network with a connection into the network. They perform no work for the defender.
At each time step the environment samples a Poisson\((\lambda)\) number of ‘‘events’’ that either activate or remove devices (while respecting a minimum network size). Newly activated non–attacker nodes are attached via Barabási–Albert preferential attachment; removed nodes simply go offline. Attacker‐owned machines remain permanently compromised and continue to influence connectivity. This dynamic topology captures the ebb and flow of hosts in a realistic enterprise network.

More precisely, at each discrete time step:
\begin{enumerate}
  \item We sample \(K\sim\mathrm{Poisson}(\lambda)\) events.
  \item Each event is independently classified as an \emph{addition} with probability \(p_{\mathrm{add}}\) or a \emph{removal} otherwise.
  \item \textbf{Addition:} select one offline node (if any remain) and add it to the network.  With probability \(p_{\mathrm{att}}\), the newly activated node is also marked \textit{attacker‐owned}; otherwise it joins as a clean host.  If its current degree in the underlying graph structure is zero (ie its never before been added), we attach it to one existing active node via preferential attachment (Barabási–Albert).
  \item \textbf{Removal:} select one active node uniformly (subject to maintaining a minimum network size) and mark it offline; its workloads and busy‐time are reset, but its record of past compromise persists.
\end{enumerate}
After each such event batch, we update the graph to reflect online/offline status and reconnect all attacker‐owned nodes to ensure they remain fully reachable from each other.  This Poisson–BA process captures the turnover of hosts in an enterprise network while preserving scale‐free connectivity and the persistence of adversary footholds. 

\subsection{Results}

\paragraph{Equilibrium vs.~Baseline Attacks and Defenses}

We consider three baselines for the defender and two for the attacker.
For both players, we compare to random attack and defense (randomly choosing among the actions at each time step), as well as ``do nothing'' baselines (no defense and no attack, respectively).
Additionally, we compare to a heuristic defense in which the defender performs a standard scanning every 7 days, and a full reset every 30 days.

Our first results compare DOAR strategies for the defender to the defense baselines, and similarly compare the attacker's DOAR strategy to several attack baselines.
These results are shown in Table \ref{tab:combined_payoffs}.
We can see that DOAR consistently outperforms all baselines for both the defender and the attacker, when the other player acts according to their DOAR strategy in both solving the POSG and a Bayesian POSG with a zero day exploit.
This is simply a confirmation that the joint strategy profile obtained by DOAR is indeed an approximate equilibrium.
What is more notable is that the defender's DOAR strategy outperforms, or performs comparably with baselines \emph{for different attack strategies} as well.
This is not self-evident, since (a) the game is not zero-sum, and (b) even in zero-sum games, robustness of equilibrium behavior does not imply that it is always a best response.
We see a similar pattern for the attacker, although DOAR attack is not a best response when the defender uses a preset heuristic policy instead of an equilibrium strategy.

%When the defender uses DOAR, the attacker’s equilibrium payoff is substantially lower than when facing either a randomly initialized network or a non‐defensive strategy. Conversely, as an attacker, DOAR secures a higher payoff against each defender strategy than a randomly generated attacker or a passive (no‐attack) policy. On the defender side, DOAR achieves a significantly higher equilibrium payoff against an adaptive attacker than a preset heuristic strategy that mirrors a common cybersecurity posture as well as a randomly generated agent. In short, DOAR drives down attacker rewards while lifting defender performance across every pairing. 

\begin{table}[htbp]
  \centering
  \caption{Average payoffs (mean\,$\pm$\,std) at equilibrium for both Attacker and Defender across defender strategies. Network configuration is available in the appendix. Results reported without the reward shaping bonus.}
  \label{tab:combined_payoffs}
  %
  % First block: Attacker payoffs
  %
  \begin{tabular}{@{} l c c c c @{}}
    \multicolumn{5}{c}{\textbf{(a) Attacker average payoffs}} \\[-0.3em]
    \toprule
    & \multicolumn{4}{c}{\textbf{Defender strategies}} \\
    \cmidrule(lr){2-5}
    \textbf{Attacker $\downarrow$ \quad Defender $\rightarrow$}
      & \textbf{DOAR} & \textbf{RandomInit} & \textbf{No Defense} & \textbf{Preset} \\
    \midrule
    \textbf{DOAR}         & $1328.000\pm0.001$  & $1429.000\pm0.001$  & $1274.000\pm0.001$  & $840.000\pm0.002$  \\
    \textbf{RandomInit} & $1085.000\pm0.001$  & $1175.800\pm39.004$ & $1236.900\pm53.217$ & $1043.900\pm43.648$ \\
    \textbf{No Attack}  & $971.000\pm0.000$   & $990.600\pm35.461$  & $917.900\pm47.359$  & $702.000\pm45.909$  \\
    \bottomrule
    \addlinespace[1em]
    %
    % Second block: Defender payoffs
    %
    \multicolumn{5}{c}{\textbf{(b) Defender average payoffs}} \\[-0.3em]
    \toprule
    & \multicolumn{4}{c}{\textbf{Defender strategies}} \\
    \cmidrule(lr){2-5}
    \textbf{Attacker $\downarrow$ \quad Defender $\rightarrow$}
      & \textbf{DOAR} & \textbf{RandomInit}  & \textbf{No Defense}  & \textbf{Preset}   \\
    \midrule
    \textbf{DOAR}         & $-14.700\pm0.001$  & $-97.367\pm0.001$  & $-38.400\pm0.002$  & $-442.217\pm0.004$ \\
    \textbf{RandomInit} & $-54.733\pm0.001$  & $-54.408\pm2.389$  & $-56.560\pm5.151$  & $-790.453\pm63.779$\\
    \textbf{No Attack}  & $-20.933\pm0.000$  & $-20.133\pm1.581$  & $-27.590\pm4.011$  & $-109.300\pm7.603$ \\
    \bottomrule
  \end{tabular}
\end{table}
%\vspace{-20pt}
\begin{table}[htbp]
  \centering
\caption{Average ex ante Bayes–Nash equilibrium payoffs (mean\,$\pm$\,std) for attacker and defender across defender strategies, with one common exploit and one private zero‐day $z\sim D_{z}$ (i.e.\ $N_{e}=1$).}
  \label{tab:combined_payoffs_bayes}
  \begin{tabular}{@{}l c c c c@{}}
    \multicolumn{5}{c}{\textbf{(a) Attacker average payoffs}} \\[-0.3em]
    \toprule
    & \multicolumn{4}{c}{\textbf{Defender strategies}} \\
    \cmidrule(lr){2-5}
    \textbf{Attacker $\downarrow$ \quad Defender $\rightarrow$}
      & \textbf{DOAR} & \textbf{RandomInit} & \textbf{No Defense} & \textbf{Preset} \\
    \midrule
    \textbf{DOAR}         & $612.000\pm0.002$ & $645.000\pm0.001$  & $954.000\pm0.001$  & $780.000\pm0.003$  \\
    \textbf{RandomInit} & $468.000\pm0.003$ & $711.638\pm78.225$ & $708.616\pm64.439$ & $539.121\pm28.427$ \\
    \textbf{No Attack}  & $240.000\pm0.050$ & $601.800\pm38.293$ & $597.900\pm43.489$ & $483.000\pm17.208$ \\
    \bottomrule
    \addlinespace[1em]
    \multicolumn{5}{c}{\textbf{(b) Defender average payoffs}} \\[-0.3em]
    \toprule
    & \multicolumn{4}{c}{\textbf{Defender strategies}} \\
    \cmidrule(lr){2-5}
    \textbf{Attacker $\downarrow$ \quad Defender $\rightarrow$}
      & \textbf{DOAR} & \textbf{RandomInit} & \textbf{No Defense} & \textbf{Preset} \\
    \midrule
    \textbf{DOAR}         & $-32.415\pm0.000$   & $-73.233\pm0.000$  & $-108.369\pm0.000$  & $-1500.085\pm0.000$ \\
    \textbf{RandomInit} & $-23.700\pm0.000$   & $-62.097\pm6.977$  & $-68.078\pm5.959$   & $-1470.797\pm9.742$ \\
    \textbf{No Attack}  & $4.620\pm0.000$     & $-63.658\pm5.829$  & $-66.671\pm5.255$   & $-1486.352\pm10.795$ \\
    \bottomrule
  \end{tabular}
\end{table}

Next, we use the framework developed to investigate the relationship between structural variables describing the nature of the organizational environment and outcomes (such as average compromise frequency) in equilibrium obtained by DOAR.

\paragraph{Impact of Relative Value of Productivity}

\begin{figure}[htbp]
  \centering
  % Scale the entire 3-panel block to 0.8×\textwidth (adjust factor as needed)
  \resizebox{0.8\textwidth}{!}{%
    \begin{minipage}{\textwidth}
      \centering
      % Panel (a)
      \begin{subfigure}[b]{0.32\textwidth}
        \centering
        \includegraphics[width=\linewidth]{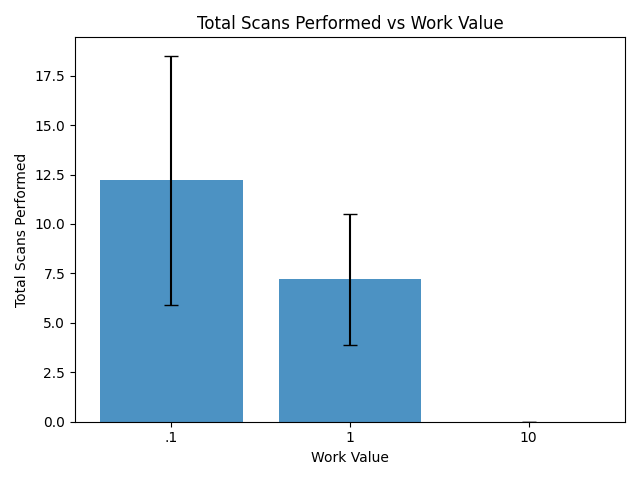}
        %\caption{Total scans performed vs.\ work value.}
        \label{fig:scans_vs_work}
      \end{subfigure}\hfill
      % Panel (b)
      \begin{subfigure}[b]{0.32\textwidth}
        \centering
        \includegraphics[width=\linewidth]{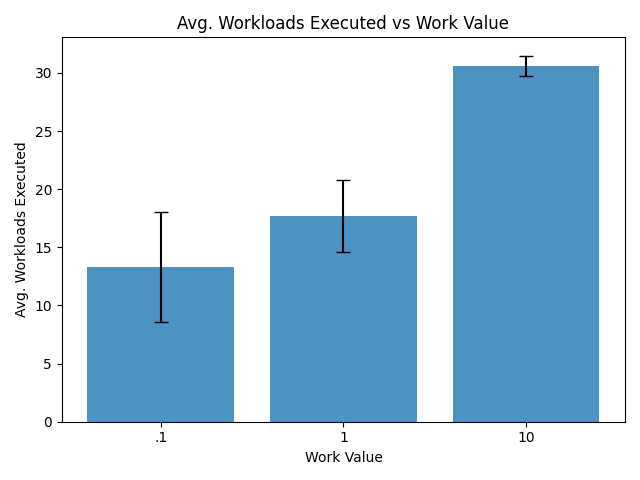}
        %\caption{Average checkpoint and reset actions per timestep vs.\ work value.}
        \label{fig:workloads_vs_work}
      \end{subfigure}\hfill
      % Panel (c)
      \begin{subfigure}[b]{0.32\textwidth}
        \centering
        \includegraphics[width=\linewidth]{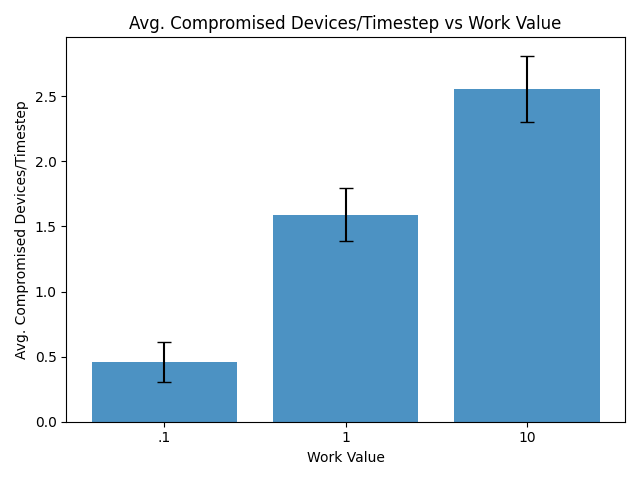}
        %\caption{Average compromised devices per timestep vs.\ work value.}
        \label{fig:compromised_vs_work}
      \end{subfigure}
    \end{minipage}%
  }
  \caption{Behavior of DOAR as a function of workload value.  
    (a) Average number of scans performed.  
    (b) Average number of workloads executed.  
    (c) Compromise rate.}
  \label{fig:work_scale_panels}
\end{figure}

We begin by considering the impact of varying the relative value of productivity as captured by \(v_w \in \{0.1,1,10\}\).  The results, provided in Figure~\ref{fig:work_scale_panels}, exhibit a tendency for the defender to abandon defense when value of work is high. Specifically, while the defender’s overall payoff (benefit minus cost) increases with \(v_w\), they perform defensive actions, such as scanning, with lower frequency, since the opportunity cost of doing so (stopping productive workflows) increases as the value of work $v_w$ rises.

\paragraph{Impact of Defense Costs}

\begin{figure}[htbp]
  \centering
  % Scale the entire 3-panel block to 0.7×\textwidth (adjust as needed)
  \resizebox{0.8\textwidth}{!}{%
    \begin{minipage}{\textwidth}
      \centering
      % Panel (a)
      \begin{subfigure}[b]{0.32\textwidth}
        \centering
        \includegraphics[width=\linewidth]{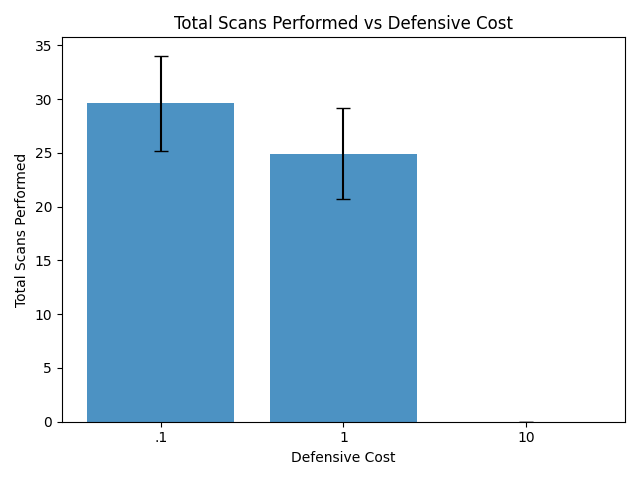}
        \label{fig:scans_vs_cost}
      \end{subfigure}\hfill
      % Panel (b)
      \begin{subfigure}[b]{0.32\textwidth}
        \centering
        \includegraphics[width=\linewidth]{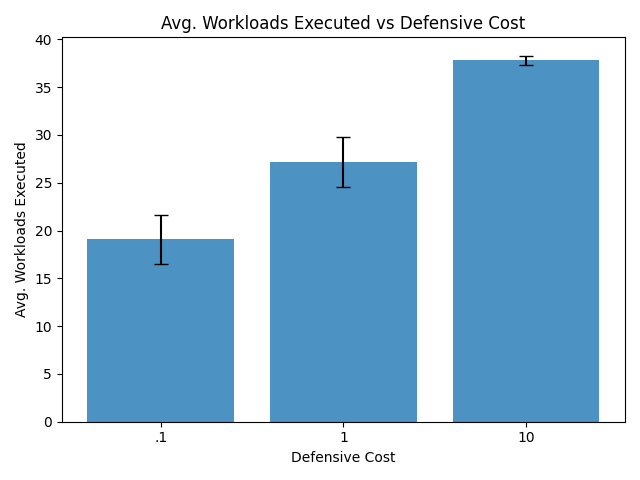}
        \label{fig:def_payoff_vs_cost}
      \end{subfigure}\hfill
      % Panel (c)
      \begin{subfigure}[b]{0.32\textwidth}
        \centering
        \includegraphics[width=\linewidth]{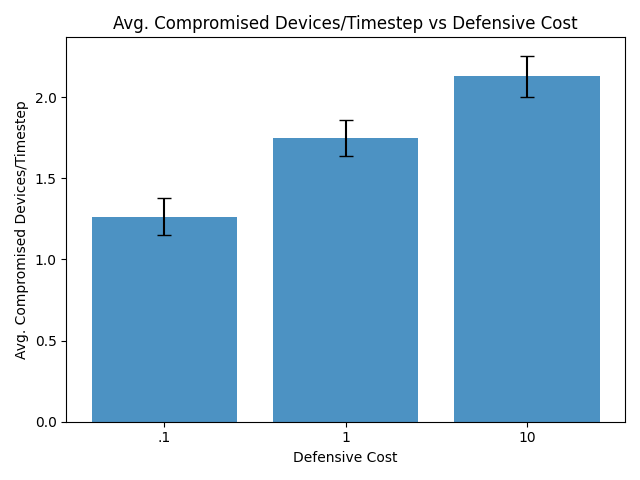}
        \label{fig:compromised_vs_cost}
      \end{subfigure}
    \end{minipage}%
  }
  \caption{Behavior of DOAR as a function of defensive costs.  
    (a) Average number of scans performed.  
    (b) Average number of workloads executed.  
    (c) Compromise rate.}
  \label{fig:cost_scale_panels}
\end{figure}

We now consider the impact of varying defensive costs. 
The results are in Figure~\ref{fig:cost_scale_panels}.
We again note a tendency of the defender to abandon defense when the cost of defense is high. In extreme cases, it abandons defense entirely. Since defensive actions stall work, we again note that as the defender defends less, the amount of work it does rises. Here, its worth noting that the advantage of defensive actions has a direct cost (the defensive cost) as well as an opportunity cost (work delayed). As either of these costs increases the defenders willingness to defend goes down.

%\vspace{-20pt}

\paragraph{Impact of System Size}

When we vary the \emph{System Network Size}, several patterns emerge in the DOAR equilibrium outcomes (Figure~\ref{fig:cost_scale_panels}). First, as network size grows, the average number of compromised devices per timestep increases.  In a small network, a few scans can quickly locate or rule out the domain controller, preventing many lateral moves.  However, in a larger topology, each individual scan covers only a small fraction of possible hosts.  Consequently, the attacker can evade detection more easily, leading to a higher compromise rate. Second, the attacker’s equilibrium payoff \emph{decreases} with larger network size.  Although compromises become more frequent, the attacker must spend more time (and possibly resources) probing a sprawling network to find the domain controller.  In effect, the “search cost” for the attacker goes up, reducing the net benefit of each successful compromise. Third, the defender’s payoff \emph{monotonically decreases} as network size increases.  With more hosts to sweep, defensive actions become less efficient at suppressing breaches, and the accumulated cost of residual compromises outweighs any scan‐savings.  Thus, the larger the network, the more negative (worse) the defender’s net payoff. Finally, we observe that the defender \emph{scans less frequently} as the network grows.  The reason is diminishing marginal returns: on a small network, each scan can drastically reduce overall compromise risk by covering a larger proportion of critical hosts (relative to the network size), but on a large network, one scan catches only a tiny slice of potential attack paths.  Since each scan still incurs a resource or time cost, the defender reduces scan frequency when facing a larger topology.  
In other words, when network size increases, scanning becomes relatively less effective at preventing lateral movement, so the equilibrium strategy calls for fewer scans despite higher compromise rates.

\begin{figure}[htbp]
  \centering
  % Scale to full text width
  \resizebox{1.0\textwidth}{!}{%
    \begin{minipage}{\textwidth}
      \centering
      % Panel (a)
      \begin{subfigure}[b]{0.24\textwidth}
        \includegraphics[width=\linewidth]{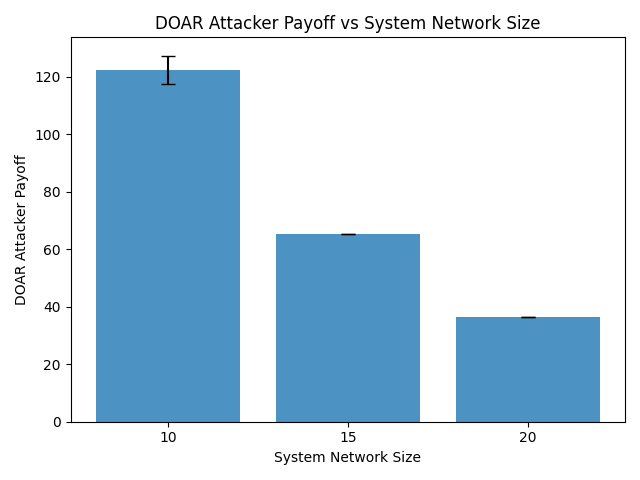}
        %\caption{Attacker payoff}
        \label{fig:network_attacker_payoff}
      \end{subfigure}\hfill
      % Panel (b)
      \begin{subfigure}[b]{0.24\textwidth}
        \includegraphics[width=\linewidth]{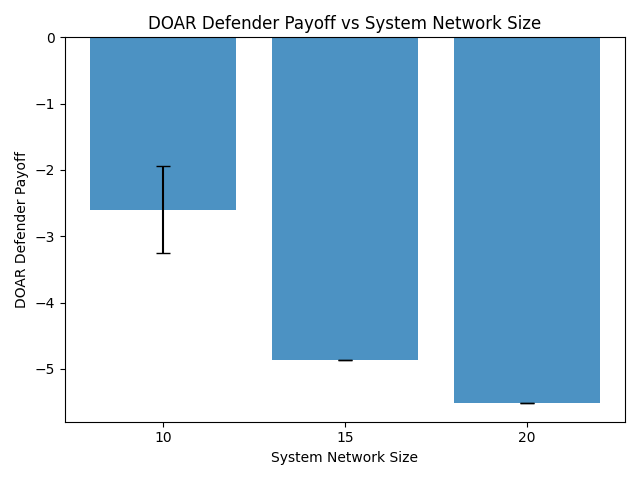}
        %\caption{Defender payoff}
        \label{fig:network_defender_payoff}
      \end{subfigure}\hfill
      % Panel (c)
      \begin{subfigure}[b]{0.24\textwidth}
        \includegraphics[width=\linewidth]{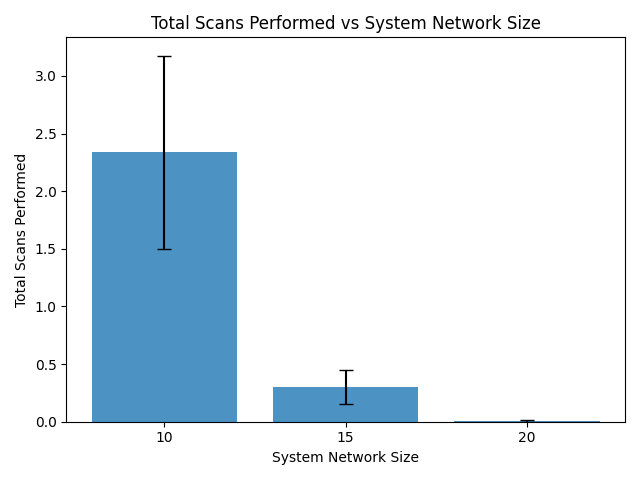}
        %\caption{Average scans}
        \label{fig:network_scans}
      \end{subfigure}\hfill
      % Panel (d)
      \begin{subfigure}[b]{0.24\textwidth}
        \includegraphics[width=\linewidth]{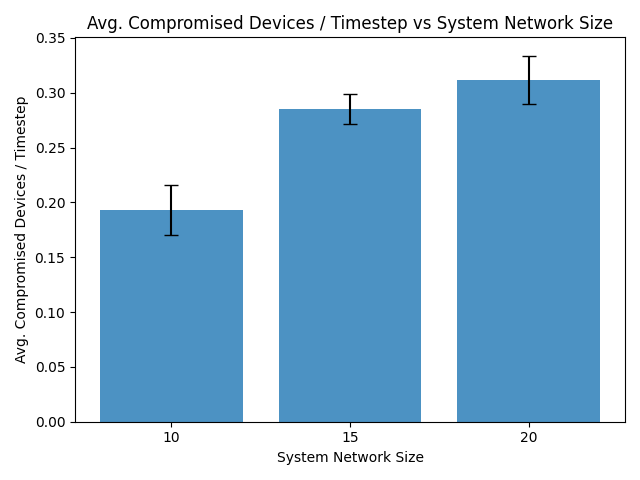}
        %\caption{Compromise rate}
        \label{fig:network_compromised}
      \end{subfigure}
    \end{minipage}%
  }
  \caption{Behavior of DOAR as a function of network size (all metrics per device).  
    (a) Attacker payoff.  
    (b) Defender payoff.  
    (c) Average number of scans.  
    (d) Compromise rate.}
  \label{fig:network_size_metrics}
\end{figure}

\paragraph{Impact of Zero Days}
Finally, we investigate the impact of zero-day exploits.
We assume here that the attacker will only have 1 additional (zero-day) exploit, but this exploit can be drawn from a distribution $D_z$.
We study, in particular, the impact of the size of $D_z$ (the number of possible zero-day options).
We consider two vulnerability‐generation regimes. In the \emph{fixed‐vulnerability} setting, the total number of zero‐day flaws is held constant: each of our 10 devices hosts up to three exploitable applications, and exactly ten zero‐day exploits are distributed across the network. In the \emph{submartingale‐vulnerability} setting, the number of flaws grows linearly with $|D_z|$, so that sampling from $D_z$ induces a submartingale in the attacker’s success probability. In both regimes, the attacker has access to a baseline exploit plus one zero‐day exploit drawn uniformly from $D_z$. We also study a variation where where the defender \emph{knows} the additional attack $z\in D_z$ sampled from $D_z$ (``known zero-day'').
This allows us to evaluate the marginal importance of the informational asymmetry inherent in zero-day attacks.
%(i.e., their posterior on that exploit becomes 1; see Section 4.4).

We observe in the \emph{fixed‐vulnerability} setting that as $|D_z|$ rises the attacker becomes less and less likely to possess a particular $z$ that can infect any particular device. Similarly, the defender becomes increasingly better as the attack can infect fewer devices. In all cases, the defender does better when it knows one of the $z\in D_z$. This advantage to the defender (and disadvantage to the attacker) is greater when $|D_z|$ is smaller, as it becomes more likely to know the particular $z$ the attacker has.

In the \emph{submartingale‐vulnerability} setting we observe the opposite behavior. With the number of compromisable devices proportional to $|D_{z}|$, the attackers payoff and compromise device count increase. The defender's payoff decreases as it performs more defensive actions trying to stop an increasingly powerful attacker.
What is particularly surprising is that even in this environment, the marginal value of knowledge of the additional attack exploit appears (at least in proportional terms) highest with fewer possible exploits available.
% Combined 2×4 grid: fixed (row 1) and submartingale (row 2)
\begin{figure}[htp]
  \centering
  % Scale to full text width (adjust factor as needed)
  \resizebox{1.0\textwidth}{!}{%
    \begin{minipage}{\textwidth}
      \centering
      % Row 1: fixed-vulnerability condition
      \begin{subfigure}[b]{0.24\textwidth}
        \includegraphics[width=\linewidth]{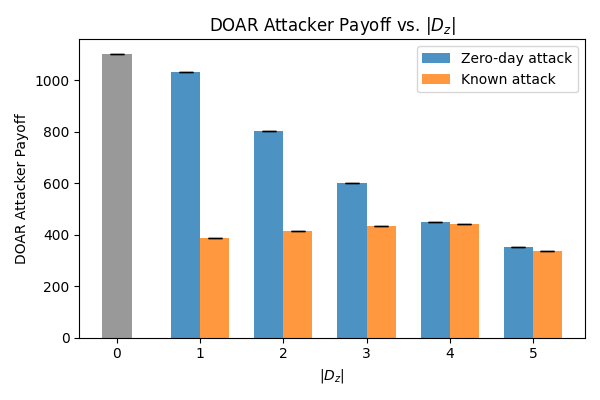}
        \caption{Attacker Payoff}
        \label{fig:do_attacker_fixed}
      \end{subfigure}\hfill
      \begin{subfigure}[b]{0.24\textwidth}
        \includegraphics[width=\linewidth]{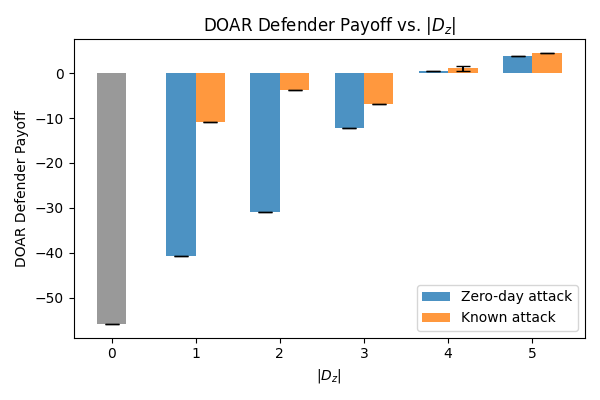}
        \caption{Defender Payoff}
        \label{fig:do_defender_fixed}
      \end{subfigure}\hfill
      \begin{subfigure}[b]{0.24\textwidth}
        \includegraphics[width=\linewidth]{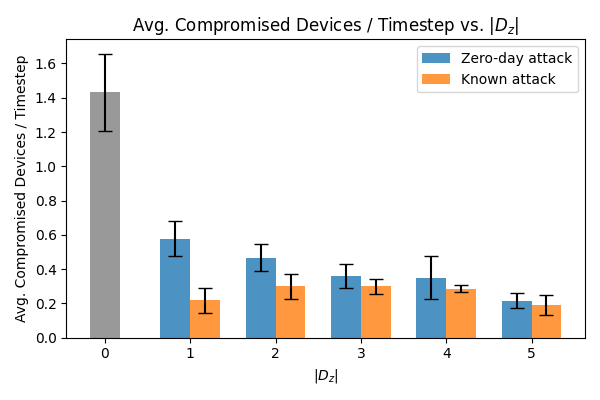}
        \caption{Compromised Devices}
        \label{fig:do_compromised_fixed}
      \end{subfigure}\hfill
      \begin{subfigure}[b]{0.24\textwidth}
        \includegraphics[width=\linewidth]{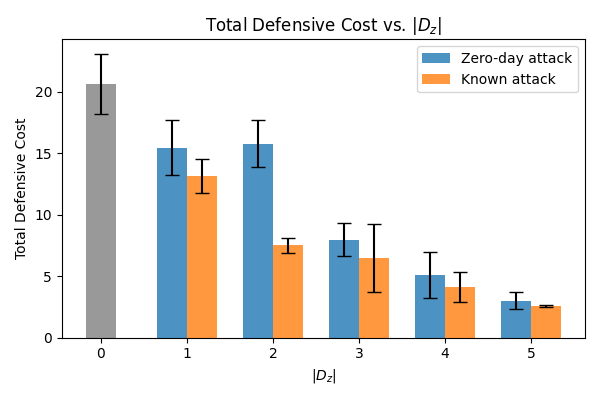}
        \caption{Defensive Cost}
        \label{fig:do_defensive_cost_fixed}
      \end{subfigure}
      \\[1ex]
      % Row 2: submartingale-vulnerability condition
      \begin{subfigure}[b]{0.24\textwidth}
        \includegraphics[width=\linewidth]{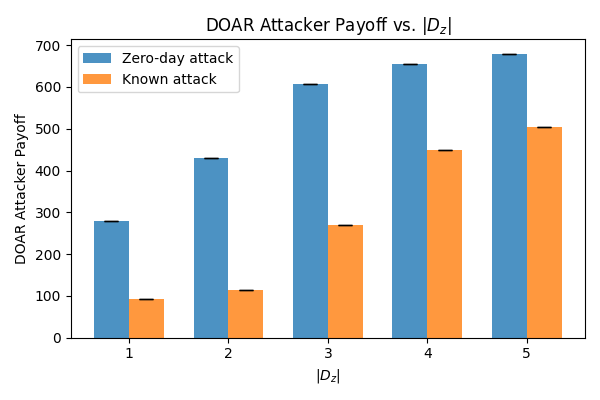}
        \caption{Attacker Payoff}
        \label{fig:do_attacker_submart}
      \end{subfigure}\hfill
      \begin{subfigure}[b]{0.24\textwidth}
        \includegraphics[width=\linewidth]{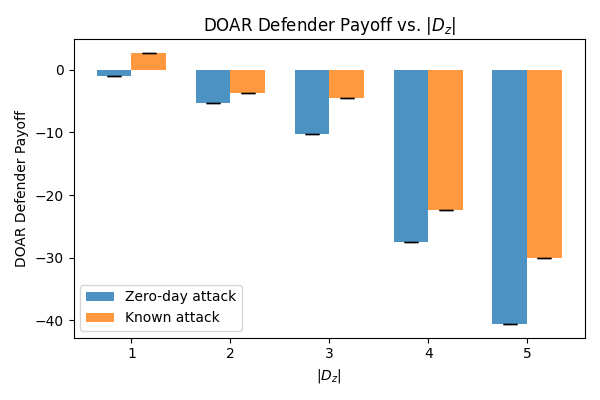}
        \caption{Defender Payoff}
        \label{fig:do_defender_submart}
      \end{subfigure}\hfill
      \begin{subfigure}[b]{0.24\textwidth}
        \includegraphics[width=\linewidth]{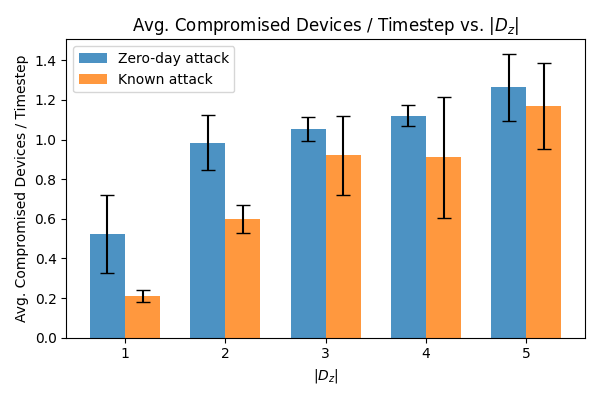}
        \caption{Compromised Devices}
        \label{fig:do_compromised_submart}
      \end{subfigure}\hfill
      \begin{subfigure}[b]{0.24\textwidth}
        \includegraphics[width=\linewidth]{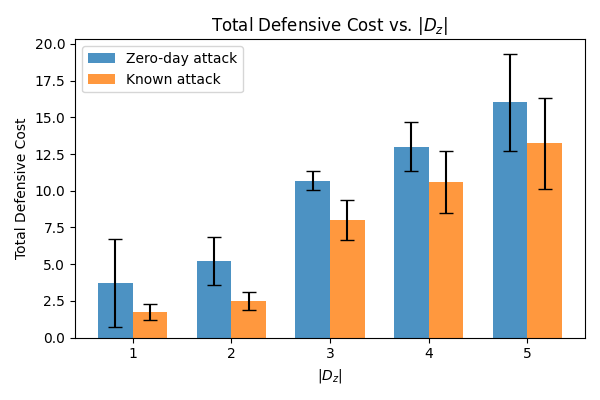}
        \caption{Defensive Cost}
        \label{fig:do_defensive_cost_submart}
      \end{subfigure}
    \end{minipage}%
  }
  \caption{Comparison of key DOAR metrics versus $|D_z|$ under both vulnerability models. Top row: fixed-vulnerability. Bottom row: submartingale-vulnerability.}
  \label{fig:do_metrics_2x4}
\end{figure}

\vspace{-25pt}
\section{Conclusion}
Herein, we have developed a comprehensive game-theoretic framework to address the challenges posed by APTs such as Volt Typhoon. By employing a combination of game theory and reinforcement learning, we created a dynamic simulation environment where defenders and attackers interact and adapt their strategies in real-time. Our custom CyberDefenseSimulator allowed us to model realistic network scenarios, incorporating diverse vulnerabilities, exploits, and defensive mechanisms.
The experimental results demonstrate that our game-theoretic approach significantly enhances the effectiveness of cyber defense strategies. The adaptive nature of the reinforcement learning agents enables them to respond intelligently to evolving threats, thereby improving network resilience. Our findings underscore the importance of proactive and strategic defensive measures in mitigating the impact of APTs on critical infrastructure.

\section{Acknowledgements}
We would like to thank Doug White and Randy Rinehart for useful industry insights.  This research was partially supported by the NSF (IIS-2214141), ONR (N00014-24-1-2663) and ARO (W911NF-25-1-0059).

\vspace{-5pt}
\section{Appendix}
% Appendix: Environment & DDPG Hyperparameters Table (compact 4‑column)
\captionsetup[table]{skip=2pt}
%\vspace{-30pt}
% Appendix: Environment & DDPG Hyperparameters Table (compact 4-column)
\begin{table}[h]
  \centering
  \caption{Environment and hyperparameters used in our experiments.}
  \label{tab:env_ddpg_params}
  \begin{tabular*}{\textwidth}{@{\extracolsep{\fill}} llll }
    \toprule
    \multicolumn{4}{l}{\textbf{Environment Parameters}} \\
    \midrule
    Devices         & 10   & Steps/Episode   & 30  \\
    MaxNetSize      & 20   & work\_scale      & 1.0 (0.01 Zero Day)  \\
    comp\_scale      & 30   & num\_attacker\_owned    & 5   \\
    Initial Compromised Ratio      & 0.4  & $\gamma$       & 0.99 \\
    def\_scale       & 1.0  &\\
    defaultversion  & 1.0  &   \\
    defaulthigh     & 3    &                 &     \\
    \midrule
    \multicolumn{4}{l}{\textbf{Best Response Hyperparameters}} \\
    reward\_scale    & 0.1  & max\_grad\_norm  & 0.5  \\
    soft-$\tau$     & 0.01 &                &      \\
    \midrule
    \multicolumn{4}{l}{\textbf{Defender Agent}} \\
    actor\_lr        & 0.001 & critic\_lr     & 0.01 \\
    buffer size      & 100k  & greedy-K       & 5    \\
    greedy-$\tau$   & 0.5   & noise\_std     & 0.1  \\
    $\lambda_e$    & 0.7   & $p\_add$         & 0.4  \\
    crit\_arch       & [158→128,128→128,128→1] &  &  \\
    \midrule
    \multicolumn{4}{l}{\textbf{Attacker Agent}} \\
    actor\_lr        & 0.001 & critic\_lr     & 0.01 \\
    buffer size      & 100k  & greedy-K       & 5    \\
    greedy-$\tau$   & 0.5   & noise\_std     & 0.1  \\
    $\lambda_e$    & 0.7   & $p\_add$         & 0.4  \\
    crit\_arch       & [111→128,128→128,128→1] &  &  \\
    \bottomrule
  \end{tabular*}
\end{table}

\newpage
\bibliographystyle{splncs04}
\bibliography{refs}
\end{document}